\documentclass[a4paper,10pt]{article}   

 \usepackage[dvips]{graphicx}
\usepackage{amssymb}
\usepackage{amsmath}
\usepackage{setspace}

\usepackage[left=3.7cm,right=3.8cm,top=5cm,bottom=5cm]{geometry}

\begin{document}
{\centerline{Accepted in "Rendiconti Lincei - Matematica et
Applicazioni (2017)"}}

{\centerline{\textbf{\emph{Chapter 41 - Mathematical Biology}}}
\vskip 0.5cm {\centerline{\textbf{\Large Continuum mechanics at
nanoscale\,: }}} \vskip 0.3cm

{\centerline{\textbf{\Large A  tool to study    trees' watering
and recovery}}}

\vskip 0.3cm

{\centerline{\textbf{Henri Gouin}}} \vskip 0.2cm
{\centerline{Aix-Marseille Univ, Centrale Marseille, CNRS, M2P2
UMR 7340, 13451 Marseille  France}} \vskip 0.2cm
{\centerline{\footnotesize{ E-mail: henri.gouin@univ-amu.fr,
henri.gouin@yahoo.fr}}}
  \vskip
0.4cm
 {\centerline{\footnotesize{\emph{\textbf{In memory of Professor Giuseppe Grioli}}}}}

\vskip 0.5cm
 {\textbf{Abstract  }}\\

 The cohesion-tension theory expounds the crude sap
ascent thanks to  the negative pressure  generated by evaporation
of water from  leaves.
 Nevertheless, trees   pose multiple  challenges and  seem
 to live in  unphysical conditions:
the negative pressure increases   cavitation;
 it  is possible to obtain a water equilibrium between connected   parts where one
  is at a positive pressure and the other  one is at negative pressure;
 no theory is    able to satisfactorily account for
  the refilling of  vessels after embolism events.

  A theoretical
 form of our paper \cite{gouinbio} in  the Journal of Theoretical Biology
  is proposed
 together with   new results:
 a continuum mechanics
 model of the disjoining pressure concept refers to the Derjaguin  school of physical chemistry.
A comparison between
 liquid  behaviour both in tight-filled  microtubes and in liquid thin-films
  is offered when the pressure  is negative  in liquid bulks and  is positive in
  liquid thin-films  and vapour bulks.
In embolized xylem microtubes,
 when the air-vapour pocket pressure is greater than the air-vapour bulk pressure, a  refilling flow occurs
 between the air-vapour  domains to empty the air-vapour pockets   although the liquid-bulk pressure remains negative.
 The model has  a limit of validity  taking the maximal size of trees into account.
  \\
These results drop an inkling that the disjoining pressure is an
efficient tool to study biological liquids in contact with
substrates at a nanoscale range.\\


\textbf{----------------------------------------------------------------------------------------------- }
\section{Introduction}

\indent

Trees are engines running on water,  but unlike animals, plants
miss an active pump to move liquids along their vascular system.
The crude sap   contains diluted salts and ascents from roots to
leaves  thanks to the water evaporation from   leaves; its
physical properties are roughly those of  water. The flow is
driven along of xylem microtubes made of dead cells which
constitute a watering network.
 Hydrodynamics, capillarity
and osmotic pressure induce the crude sap ascent of a few tens of
meters only \cite{Zimm}; nevertheless
    a sequoia   of 115.55 meters height   is living in California \cite{Preston1}.
    Additively, trees operate a second vascular system
- phloem sieve tubes - for the circulation of metabolites through
their
 living tissues and elaborated sap flows passing from leaves to roots.
   \\
 Measurements
of the pressure within the terminal xylem vessels illustrate an
extraordinary consequence of the tree behaviour for moving water:
the liquid water is under tension. An   experimental checking
comes from an apparatus called Scholander pressure chamber
 (see Fig.\ref{fig1}, \cite{Scholander}).
 \begin{figure}[h]
\begin{center}
\includegraphics[width=8cm]{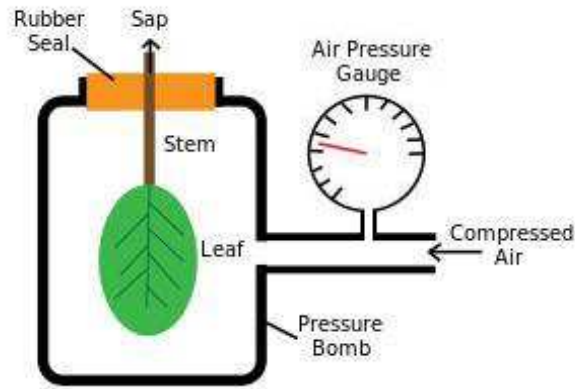}
\end{center}
 \caption{ \footnotesize {Sketch of the Scholander pressure chamber}.
  A leaf attached to a stem is placed inside a sealed chamber.
  Compressed air is  slowly added to the chamber. As the pressure
  increases to a convenient level, the sap is forced out of the xylem
   and is visible at the cut end of the stem. The   required pressure is
    opposite and of equal magnitude to the water pressure in the water-storing
     tracheids in the leaf.} \label{fig1}
\end{figure}
The pressure difference across plants can easily be of the order
of 1 to 10 MPa \cite{Meyra}. Although trees do not approach the
ultimate tensile strength of liquid water during transpiration
\cite{Herbert}, multiple types of measurements provide evidence
for cavitation of liquid water  in the xylem microtubes and
cavitation events   have been acoustically detected with
ultrasonic transducers pressed against the external surface of
trees \cite{Milburn,Tyree}. The porous vessel walls can prevent
the  bubbles from spreading \cite{Mercury} and the principal flow
of water during transpiration goes to evaporation through stomata
on the underside of leaves. The pores - or bordered pits -
connecting adjacent segments in the xylem vessels pass through the
vessel walls, and are bifurcated by bordered-pit membranes which
are thin physical fluid-transmitters. No vessels are continuous
from roots  to petioles, and the water does not leave the vessels
in the axial direction  but laterally along a long stretch
\cite{0'Brien}.  When wetted on both sides, the bordered-pit
membranes allow the liquid-water flow to pass through.   In the
leaves, the membranes serve as capillary seals; in the stems, the
bordered-pit membranes also serve as seals between a gas-filled
segment and an adjacent liquid-filled segment avoiding propagation
of massive embolisms \cite{Tyree1}. Consequently, trees seem to
live in unphysical conditions \cite{Zwieniecki}; to be hydrated,
they exploit liquid water in metastable states at negative
pressure \cite{Holbrook}.

A classical explanation of the sap ascent phenomenon in tall trees
is the  \emph{cohesion-tension theory} propounded in 1893-1895 by
Boehm, Dixon and Joly  and Askenasy \cite{Askenasy,Boehm,Dixon},
followed by an analysis of the sap motion propounded by van der
Honert \cite{vanderHonert}. According to this theory, the crude
sap tightly fills microtubes of dead xylem cells and its transport
is due to a gradient of negative pressure producing the traction
necessary to lift water against gravity. The decrease in  negative
pressure is related to the closing of  aperture of microscopic
stomata in leaves through which water vapour is lost by
transpiration.
 The considered aperture
  is about 2 $\mu$m - or less at the top of tall trees as
suggested in  \cite{Koch,Pridgeon} - which is the right size to
prevent cavitation for nucleus germs of the same order of
magnitude. Nonetheless, several objections question the validity
of the cohesion-tension theory, and worse,   preclude the
possibility of refilling embolized xylem tubes. To this goal, we
first refer to the textbook by Zimmermann \cite{Zimm}. He said:
\\\emph{The heartwood is referred to as a wet wood. It may
contain liquid under positive pressure while in the sapwood the
transpiration stream moves along a gradient of negative pressures.
Why is the water of the central wet core not drawn into the
sapwood? The heartwood is relatively dry i.e. most tracheids are
embolized. It is rather ironic that a wound in the wet wood area,
which bleeds liquid for a long period of time, thus appears to
have the transpiration stream as a source of water, in spite of
the fact that the pressure of the transpiration stream is negative
most of the time. It should be quite clear by now that a drop in
xylem pressure below a critical level causes cavitations and
normally puts the xylem out of function permanently.}\\ At great
elevation, the value of the negative pressure increases   risks of
cavitation and consequently, the formation of embolisms may cause
a definitive break-down of  continuous columns of sap inducing
leaf death. Crude sap is a fluid with   superficial tension
 lower than   superficial tension $\gamma$ of the pure water,
  which is about $72.5 \times 10^3
$ N.m$^{-1}$ at $20^\texttt{o}$ Celsius \cite{Meyra}; if we
consider a microscopic air-vapour bubble with a diameter $2R$
smaller than xylem microtube diameters, the difference between
air-vapour pressure $\wp_v$ and liquid sap pressure $\wp_l$ is
expressed by   Young-Laplace formula $\wp_v - \wp_l = 2\gamma /R$;
the air-vapour pressure is positive and consequently unstable
bubbles will appear when $R \geq - 2\gamma/\wp_l$. For a negative
pressure $\wp_l =  - 0.6$ MPa in the sap, corresponding to an
approximative minimal value of the hydrostatic pressure for
embolism reversal in plants of Laurus nobilis \cite{Nardini}, we
obtain $R \geq 0.24\, \mu$m; then, when all the vessels are
tight-filled, germs naturally
pre-existing in crude water may spontaneously embolize the tracheids.\\
Another objection to the confidence in the cohesion-tension theory
was also the experiment which demonstrated that tall trees
survived double saw-cuts, made through the cross-sectional area of
the trunk to sever all xylem elements, by overlapping them
\cite{Preston}. This result, confirmed by several authors does not
seem to be in agreement with the possibility of strong negative
pressures in the water-tight microtubes \cite{Mackay,Sperry}.
Using a xylem pressure probe, Balling \& Zimmermann \cite{Balling}
showed that, in many circumstances, the apparatus does not measure
any water tension \cite{Tyree}. However, there are other
possibilities for the tree survival and researchers presented
some experimental evidences for the local refilling that restores
embolized conduits by visualizing the conduits with microscope
\cite{Canny0,Canny1,Holbrook1,McCully}.
\\
A negative argument also seems to come from the crude sap
recovery in embolized xylem tubes.  At high elevation, it does not
seem possible to refill a tube full of air-vapour at a positive
pressure when liquid-water is at a negative pressure. In xylem,
the liquid-water  metastability - due to negative pressures - may
persist even in the absence of transpiration. Consequently,
refilling processes pose a tough physical challenge to push the
liquid-water back into  xylem vessels: once embolized vessels have
reached a nearly full state, is the refilling solution still at
positive pressure  with some remaining air?
\\
The most popular theory of refilling process has been proposed in
several papers. Due to the fact that xylem microtubes are
generally in contact with numerous living cells \cite{Zimm}, it is
hypothesized that crude sap is released into the vessel lumen from
the adjacent living cells in a manner similar to root exudation
\cite{Kramer} and it is assumed that the mechanism for water
movement into embolized conduits involves the active secretion of
solutes by the living cells \cite{Lampinen}. Nonetheless, a survey
across species indicated that the root pressure can reach 0.1-0.2
MPa above atmospheric pressure \cite{Fisher} and seems the only
logical source of embolized vessels
 repairing
at night in smaller species with well-hydrated soil. The  Munch
pumping mechanism \cite{Munch} was invoked, but basic challenges  still persist: osmotic pressures measured
in sieve tubes do not scale with the height of a plant as one
would expect \cite{Turgeon} and such scenarios have not yet been
empirically verified. Hydraulic isolation is also required to
permit the local creation of the positive pressures necessary to
force the gas into solution and the embolism removal may be
concurrent with tree transpiration \cite{Zwieniecki1}. Additively, refilling in the presence of tension in
adjacent vessels requires the induction of an energy-dissipating
process that locally pumps liquid into the emptied vessels \cite{Canny0} or lowers the water potential
in the vessel with the
secretion of solutes \cite{Zwieniecki}. As a
consequence, many authors suggested that
alternative mechanisms must be required \cite{Johnson,Zimmu}.\\

Nowadays,
 the  development of techniques allows us to observe phenomena at
length scales of a very few number of nanometers. This
nanomechanics reveals new behaviors, often surprising and
essentially different from those that are usually observed at
macroscopic but also at microscopic scales \cite{Bhushan,Garajeu}.
As pointed out in experiments, the density of  water is found to
be changed in narrow pores. The first reliable evidence of this
effect was reported by B.V. Derjaguin, V.V. Karasev  and E.N.
Efremova \cite{Karasev} and found after by many others
\cite{Derjaguin}, pp. 240-244. In order to evaluate the structure
of thin interlayers of water  and other liquids, Green-Kelly and
Derjaguin employed a method based on measuring changes in
birefringence \cite{Green-Kelly} and they found significant
anisotropy of  interlayers.
\\
Slightly compressible liquids   wetting solid substrates  point
out an unexpected behaviour in which liquids do not transmit the
pressure to all their connected domains \cite{Lifshitz}:
  it is possible to
obtain an equilibrium between connected  parts where one is at
positive pressure - the pressure in a liquid thin-film - and the
other is at  negative pressure - the pressure in the liquid bulk.
The air-vapour phase in contact with the liquid thin-film is at
the same positive pressure as the liquid thin-film. The
experiments and model associated with this behaviour fit the
disjoining pressure concept  \cite{deGennes,Derjaguin} which is a
well adapted tool for a very thin liquid film of thickness $h$. In
cases of Lifshitz' analysis \cite{Lifshitz} and van
der Waals' theory \cite{vdW}, behaviours of disjoining pressure $\Pi$  are respectively as $%
\Pi\sim h^{-3}$ and $\Pi\sim \exp ( -h)$. None of them fits
 experimental results for a film with a thickness ranging
over a few nanometers.
\\
 Since van der Waals, the fluid
inhomogeneities in liquid-vapour interfaces have been represented
with continuous models by taking a volume energy depending on
space density derivative  into account
\cite{Isola,Slemrod,kaz,seppecher,Widom}. Nevertheless, the
corresponding square-gradient functional is unable to model
repulsive force contributions and misses the dominant damped
oscillatory packing structure of liquid interlayers near a
substrate wall \cite{chernov2,Weiss}. The decay lengths are
correct only close to the liquid-vapour
critical point where the damped oscillatory structure is subdominant \cite%
{Evans1,Evans3}.  In contrast, fluctuations strongly damp
oscillatory structure and it is mainly for this reason that van
der Waals' original prediction of a hyperbolic tangent curve in density is so close
to simulations and experiments \cite{Rowlinson}.
\\
To propose an analytic expression in density-functional theory for
liquid film of a very few nanometer thickness near a solid wall,
we add a liquid energy-functional at the solid surface and a
surface energy-functional at the liquid-vapour interface to the
square-gradient functional representing the volume free energy of
the fluid. This kind of functional is well-known in the literature
\cite{Fisher0} and the process is simpler than the renormalization
group theory \cite{Fisher1,Gold} mainly used near critical points.
It was used by Cahn in a phenomenological
form  \cite%
{Cahn0}. An asymptotic expression is obtained in \cite{gouin} with
an approximation of hard sphere molecules for liquid-liquid and
solid-liquid interactions: in this way, we also took   account of
the power-law behavior which is dominant in a thin liquid film in
contact with a solid  \cite{Israel}.\\

\emph{ The paper is organized as follows:}
 Section 2 expounds   that nanofluidic and liquid thin-films concepts are  fundamental  tools used in the paper; following Derjaguin's Russian school of physical chemistry, we  propose an experimental overview of the disjoining pressure concept for liquid thin-films at equilibrium.
Section 3 is an analytical and numerical study of the disjoining pressure along vertical
  liquid thin-films. Section 4 studies the liquid motions along vertical liquid thin-films, and
Section 5 is a comparison between
 liquid-motions' behaviours both in tight-filled  microtubes and in liquid
 thin-films.
Section 6 focuses on trees containing  vessels considered as machines. From experiments presented in  previous sections, a model of xylem    using liquid thin-films is proposed.
   Such a    model of xylem allows to explain both the thermodynamical  consistence of
   the cohesion-tension theory and the conditions of the crude-sap refilling     at high elevation.
This previous \emph{thought experiment}   is modified to take  account of air-vapour pockets:
 when the air-vapour pocket pressure is greater than the air-vapour bulk pressure, a huge flow occurs
 between the two parts filled by air-vapour gas to empty the air-vapour pockets although the liquid-bulk pressure is negative. Finally, the   \emph{pancake-layer concept}, associated with the
breaking-down of vertical liquid thin-films, allows to forecast
the limit of validity of the model and yields a maximum   height
for the tallest trees.   \newline We present new results
concerning models and numerical calculations for comparing filled
microtube  motions  and   thin-film motions, a new study of
laterally transfer of masses between xylem microtubes and an
explanation for ultrasounds eventually generated in the watering network. \\
  A conclusion ends
the article. The apparent incompatibility between the model in
\cite{Gouin8} and the cohesion-tension theory is now solved.
Experiments are suggested to  verify the accuracy of the sap
ascent for tall trees and   of the crude-sap's refilling.

\section{The disjoining pressure}

\noindent The disjoining pressure   concept is associated with
liquid thin-films bordered by vapour  bulks and wetting    flat
solid surfaces. Experiments and analysis are described by
Derjaguin \emph{et al} \cite{Derjaguin}. At    {given temperature
$T_{_0}$, two experiments allow to understand  the physical
meaning of horizontal  liquid thin-films at equilibrium.

 $\bullet$ The first experiment   was  described in \cite{Derjaguin} pp. 330--331:   a liquid
bulk   submitted to  pressure $P_{l_{b}}$
contains a microscopic bubble of radius $R$ contiguous to a solid
 (Fig. \ref{fig2}). The  bubble floats upward and approaches a horizontal smooth   plate, and a planar liquid thin-film is formed after some time.
The liquid thin-film separates the flat part of the bubble which
is squeezed onto
 the solid surface, from inside. Inside the bubble,  the pressure of   vapour bulk  $ {v_{b}} $ of density
$\rho_{v_{b}}$ (\emph{mother vapour-bulk})  is $P_{v_{b}}$. The
film is thin enough for gravity to be neglected thickness-wise and
the hydrostatic pressure of the liquid thin-film  is identical to
the vapour-bulk pressure inside the bubble. Pressure $P_{v_{b}}$
differs from  pressure $P_{l_{b}}$ of liquid bulk   $ l_{b} $ of
density $\rho_{l_{b}}$ (\emph{mother liquid-bulk})
\cite{Derjaguin}, page 32. The    analysis can apply to the bulk
pressure $P_{l_{b}}$ in the liquid at  short distance away from
the surface;
 bulk pressure $P_{l_{b}}$ is not really affected by the
gravity because of the microscopic size of the bubble which
remains spherical outside the liquid thin-film.  The Young-Laplace
formula describes the difference between the two bulk pressures:
\begin{figure}
\begin{center}
\includegraphics[width=9cm]{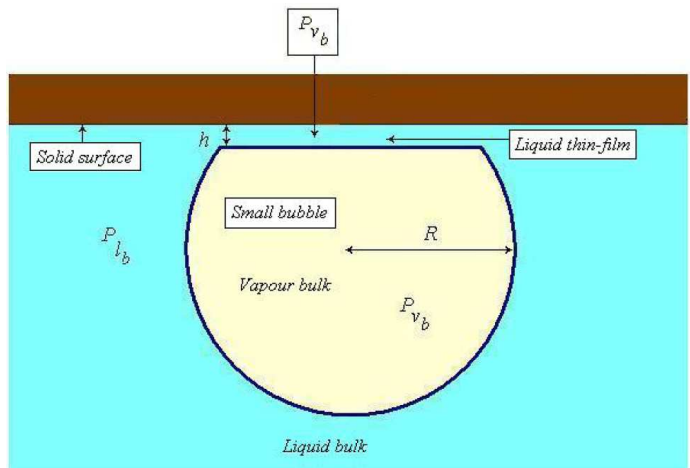}
\end{center}
\caption{\footnotesize {The bubble method of determining the
disjoining pressure isotherms of wetting films.  The hydrostatic
pressure in the liquid thin-film is the same as in the microscopic
bubble  and is different from the liquid-bulk pressure.}}
\label{fig2}
\end{figure}
\begin{equation}
P_{v_{b}}- P_{l_{b}} = \frac{2\gamma}{R},\label{Laplace}
\end{equation}
 where $\gamma$ is the surface tension   of the bubble liquid-vapour interface.
The liquid thin-film extends from the bulks which create  the
pressure difference already estimated in Eq. \eqref{Laplace}  and
named $\Pi(h)$:
\begin{equation}
\Pi(h)  =P_{v_{b}}-P_{l_{b}}\, .  \label{disjoiningpressure}
\end{equation}
Interlayer   pressure  $\Pi(h) $ additional to the mother liquid-bulk pressure is called
 the disjoining pressure of the thin film of thickness $h$, and   curve\
 $h \longrightarrow\Pi(h)$\ \ - obtained by changing the bubble's radius   and thereby   film thickness $h$ -  is the
\emph{disjoining pressure isotherm}.
\\
$\bullet$   The second experiment  is associated with
the    apparatus due to Sheludko
 \cite{Sheludko} and   described in Fig. \ref{fig3}.
The film is thin enough such that the gravity effect is neglected
across the liquid layer. The hydrostatic pressure in the thin
liquid layer included between a solid wall and the vapour bulk
differs from the pressure in the contiguous liquid bulk from which
the liquid layer extends (this is the reason
  for which Derjaguin used  mother-bulk term). The forces arising during the
thinning of the film of uniform thickness $h$ produce the
disjoining pressure which is the additional pressure on the
surface of the film
 to the pressure within the  mother liquid-bulk.
  Clearly, a disjoining pressure could
be measured by applying an external pressure to keep the complete
layer in equilibrium and   verifies Eq.
\eqref{disjoiningpressure}.
\begin{figure}[h]
\begin{center}
\includegraphics[width=12cm]{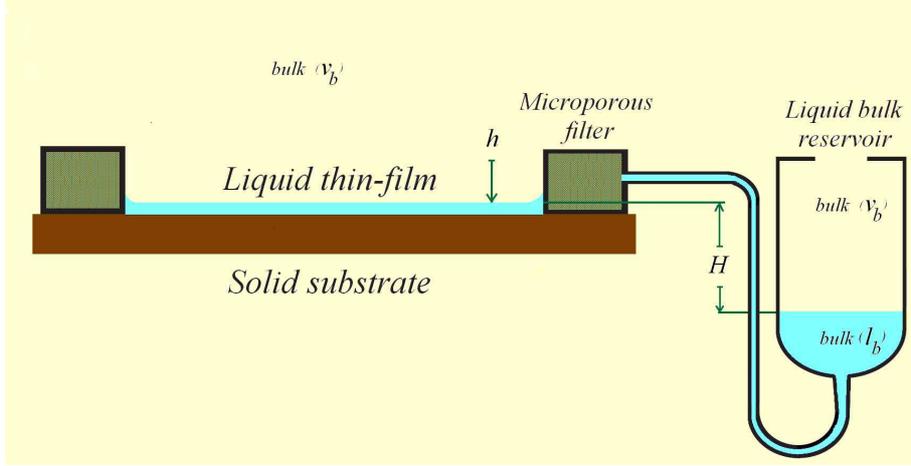}
\end{center}
\caption{\footnotesize{Diagram of the technique for determining
the disjoining pressure isotherms of wetting films on a solid
substrate: a  circular wetting film is formed on a flat substrate
to which a microporous filter is clamped. A pipe connects the
filter filled with the liquid to a reservoir containing the mother
liquid-bulk that can be moved by a micrometric device. Thickness
$h$ of the film depends on   $H$   in a convenient domain of
$H$-values, where the wetting film is stable. The disjoining
pressure is equal to $\Pi =
(\rho_{b}-\rho_{v_b})\,\mathcal{G}\,H$, where $\mathcal{G}$ is the
acceleration of gravity  (\cite {Derjaguin}, page
332).}}\label{fig3}
\end{figure}

Derjaguin's clever idea was to create an analogy between liquid thin-films and liquid-vapour
interfaces of bubbles. Liquid thin-films - allowing  to obtain an equilibrium between
fluid phases   at different pressures - are physically similar to  \emph{bubbles' flat interfaces}.
 The pressure in the liquid phase is different  from the liquid pressure in the liquid thin-film,
  which is the same as  the pressure in the vapour phase; thereby, the liquid
 does not completely transmit  the  pressure in all places where it
 lays.

Let us consider the Gibbs free energy  of the  liquid layer
(thermodynamic potential). As pointed out by Derjaguin \emph{et al} in
 (\cite{Derjaguin}, {Chapter 2}), at temperature $T_0$,
 the   Gibbs free energy per unit area $G$ can be simply expressed as a function of
 $h$:
\begin{equation*}
\frac{dG(h)}{dh} = -\Pi(h), \label{Gibbs1}
\end{equation*}
and   can be integrated as:
\begin{equation}
G (h) = \int_{h}^{+\infty} \Pi(h)\,dh,\label{Gibbs2}
\end{equation}
where $h=0$ is associated with the dry wall in contact with the
vapour bulk and $h=+\infty$ is associated with a wall in
 contact with   liquid bulk  when the value of $G$ is $0$.\\
 An important property related to the  problem of wetting
 is associated with the  spreading coefficient \cite{degennes2}:
 \begin{equation*}
S = \gamma_{_{SV}} - \gamma_{_{SL}}-\gamma_{_{LV}},\label{wetting}
\end{equation*}
where  $\gamma_{_{SV}}, \gamma_{_{SL}}, \gamma_{_{LV}}$ are
respectively the  solid-vapour, solid-liquid and liquid-vapour
free energies per unit area of interfaces.
 The liquid-layer  energy per unit area can be written as
 \begin{equation*}
E =  \gamma_{_{SL}}+\gamma_{_{LV}}+ G(h). \label{layer energy}
\end{equation*}
When $h=0$, we obtain the energy $\gamma_{_{SV}}$ of the dry solid
wall; when $h=+\infty$, we obtain $\gamma_{_{SL}} +
\gamma_{_{LV}}$. In complete wetting of liquid on  solid wall, the
spreading coefficient is positive.
\\
The conditions of stability of a liquid thin-film essentially
depend on phases between which the film is sandwiched. In case of
 single film in equilibrium between the vapour and a solid
substrate, the stability condition is classically:
 \begin{equation*}
 \frac{\partial \Pi (h)}{{\partial h}} < 0\quad\Longleftrightarrow\quad
 \frac{\partial^2 G (h)}{{\partial h^2}} > 0 .\label{stability}
 \end{equation*}
\begin{figure}[h]
\begin{center}
\includegraphics[width=9cm]{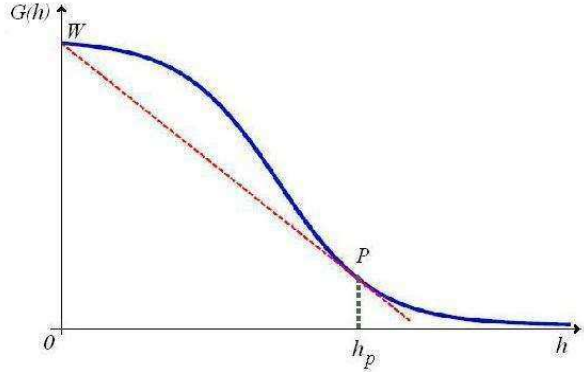}
\end{center}
\caption{  \footnotesize The graph is  the sketch of  Gibbs
free-energy G(h) for liquid water  in contact with a wetting solid
wall. The construction of the tangent to  curve $G(h)$ issued from
point
 $W$  of coordinates  $( 0,G(0))$  yields point  $P$; point
$W$ is associated with a high-energy surface  of the dry wall and point   $P$  is associated with
 pancake-thickness  $h_p$  \cite{degennes2}.} \label{fig4}
\end{figure}
The coexistence of two film segments with different thicknesses is a phenomenon which can be interpreted with
 the equality of chemical potential and superficial tension of the two films. A spectacular case
  corresponds to the coexistence of a liquid film of thickness $h_p$ and the dry solid wall
  associated with $h=0$.  The film is the so-called  \emph{pancake  layer}  corresponding to
  condition
  \begin{equation}
G(0) =  G(h_p)+ h_p\, \Pi(h_p). \label{pancake thickness}
\end{equation}
 Equation (\ref{pancake thickness}) expresses that the value of the Legendre transformation of $G(h)$ at $h_p$ is equal to $G(0)$.
 Liquid films of thickness $h > h_p$ are stable and liquid films of thickness $h < h_p$ are metastable or unstable. Thickness $h_p$ can be obtained by the geometric transformation  drawn  in Fig.
\ref{fig4}.

\section{Equilibrium of vertical liquid thin-films}

\subsection {A functional of energy associated with liquid thin-films}

The modern understanding of liquid-vapour interfaces begins with
papers of van der Waals \cite{Ono,vdW}. In current approaches, it
is possible to give exact expressions of the free energy in terms
of pair-distribution function and   direct correlation function
\cite{Lutsko}. In practice, these complex expressions must be
approximated to lead to a compromise between accuracy and
simplicity.
 When we are confronted with such complications, the mean-field models are generally
 inadequate and the obtained qualitative picture is no more sufficient. The main alternatives are
 density-functional theories which are a lot simpler than the Ornstein-Zernike equation
 in statistical mechanics since the local density is a functional at each point of the fluid
  \cite{Evans1,Rowlinson}.
  We use this approximation enabling us to  analytically compute the density profiles of simple fluids
  where we take  account of surface effects and repulsive forces by adding density-functionals at boundary surfaces.
The energy  functional of the inhomogeneous fluid in a domain $O$
of boundary $\partial O$  is taken in the form:
\begin{equation*}
F_o = \iiint_O \rho\,\varepsilon\ dv + \iint_{\partial O} \varpi\
ds .\label{density functional}
\end{equation*}
 The first integral is associated with a square-gradient approximation when we introduce a specific
free energy $\varepsilon$ of the fluid at a given temperature
$T_0$,   as a function of density $\rho$ and  $\beta= ({\rm
grad}\, \rho)^2$ and $\varpi$ is a convenient surface energy.
Specific free energy $\varepsilon$ characterizes together the
fluid properties of compressibility and capillarity.
 In accordance  with   kinetic
theory,
\begin{equation*}
\rho\, \varepsilon =\rho\, \alpha (\rho)+\frac{\lambda
}{2}\,({\rm{grad\, }}\rho )^{2},  \label{internal energy}
\end{equation*}
where term $ ({\lambda }/{2})\,(%
\rm{grad\ \rho )^{2}}$  is added  to the volume free-energy $\rho
\,\alpha (\rho)$ of the  compressible-fluid bulk and coefficient
$\lambda =2\rho\, \varepsilon _{\beta }^{\prime }(\rho, \beta)$\
(\,' denotes the partial derivative) is assumed to be constant at
given temperature $T_0$ \cite{Rocard}. Specific free energy
$\alpha $ enables to connect continuously liquid and vapour bulks,
and thermodynamical
 pressure $P(\rho)=\rho
^{2}\alpha _{\rho }^{\prime }(\rho )$ is a state equation for
liquid-vapour interfaces like  van der Waals' pressure.
\newline
Near a solid wall, London potentials of liquid-liquid and
liquid-solid interactions are
\begin{equation*}
\left\{
\begin{array}{c}
\displaystyle \;\varphi _{ll}=- {c_{ll}}/{r^{6}}\;,\text{ \ when\
}r>\sigma _{l}\;\;\text{and }\;\ \varphi _{ll}=\infty \text{ \
when \ }r\leq
\sigma _{l}\, ,\  \\
\displaystyle \;\varphi _{ls}=- {c_{ls}}/{r^{6}}\;,\text{ \ when\
}r>\delta \;\;\text{and }\;\ \varphi _{ls}=\infty \text{ \ when \
}r\leq \delta \;, \
\end{array}
\right.
\end{equation*}
where $c_{ll}$ and $c_{ls}$ are two positive constants associated
with Hamaker coefficients, $\sigma _{l}$ and $\sigma _{s}$ denote
fluid and solid  molecular-diameters, $\delta =\frac{1}{2}($
$\sigma _{l}+$ $\sigma _{s})$ is the minimal distance between
centers of fluid and solid molecules \cite{Israel}. Forces between
liquid and solid  are short range and can be simply described by
adding a special energy at the surface. This energy is the
contribution to the solid/fluid interfacial energy which comes
from direct contact.
 For a plane solid wall  at a molecular scale, this
  surface free energy is in the form:
\begin{equation}
\phi(\rho)=-\gamma _{1}\rho+\frac{1}{2}\,\gamma_{2}\,\rho^{2}.
\label{surfaceenergy}
\end{equation}
Here $\rho$ denotes the fluid density value  at surface $(S)$;
constants $\gamma _{1}$, $\gamma _{2}$ and $\lambda$ are positive
and given by the mean field approximation:
\begin{equation*}
 \gamma
_{1}=\frac{\pi c_{ls}}{12\delta ^{2}m_{l}m_{s}}\;\rho _{sol},\quad
 \gamma _{2}=\frac{\pi c_{ll}}{12\delta^2 m_{l}^{2}},\quad  \lambda =   \frac{2\pi
c_{ll}}{3\sigma_l \,m_{l}^{2}},\label{coefficients}
\end{equation*}
where $m_{l}$ and $m_{s}$ denote  masses of fluid and solid
molecules,   $\rho _{sol}$ is the solid density
\cite{deGennes,gouin}.

This is not  the entire interfacial energy: another contribution
 comes from the distortions in the density profile near the wall \cite{Cahn0,deGennes,gouin}.
We consider a   liquid layer contiguous to its vapour bulk and in
contact with a plane solid wall $(S)$; the z-axis is perpendicular
to the solid surface.
 The
conditions in  the vapour bulk are  $\displaystyle {\rm grad}\,
\rho =0$ and
 $\Delta \rho = 0$ where $ \Delta$ denotes the Laplace operator.\\
 Far below from the fluid critical-point,
 a way to compute the total free energy of the complete liquid-vapour
 layer is to add the surface energy of  solid wall $(S)$ at $z=0$,
  the energy of  liquid layer $(L)$
 located between $z=0$ and $z=h$, the energy of the sharp liquid-vapour
 interface of a few Angstr\"{o}m thickness assimilated to surface $(\Sigma)$
  at $z=h$ and the energy of the vapour layer located between $z=h$ and
$z=+\infty$ \cite{Gavrilyuk}. The liquid at level $z=h$ is
situated at a distance order of two molecular diameters from the
vapour bulk and the vapour has a negligible density
 with respect to the liquid density \cite{Pismen}. \\
 In our model, the two last energies can be expressed with writing a unique
  energy $\psi$ per unit surface located on    mathematical surface $(\Sigma)$ at  $z=h $:
by a calculus like in  \cite{gouin}, we can write $\psi$ in the
same form than Rel.\,\eqref{surfaceenergy}  and   expressed in
\cite{deGennes} like $\psi(\rho) = -\gamma
_{5}\rho+({1}/{2})\,\gamma_{4}\,\rho^{2}$; but vapour density
being negligible, $\gamma _{5}\simeq 0$ and  surface free energy
$\psi$ is reduced to
\begin{equation}
\psi (\rho )=\frac{\gamma _{4}}{2}\ \rho_h ^{2}, \label{cl2}
\end{equation}
where $\rho_h$ is the liquid density at level $z=h$ and  $
\gamma_4$ is associated with   distance $d$ of the  order of the
fluid molecular diameter (then $d \simeq  \delta$ and
$\gamma_4\simeq \gamma_2 $).

The properties of Section 2 can be extended to vertical thin-films
when we add   the gravitational potential $\Omega $ to the energy
functional; we obtain a functional $F$ in final form
\begin{equation}
F = \iiint_{(L)}\rho\, \varepsilon\ dv +\iiint_{(L)}\rho\, \Omega
\ dv + \iint_{(S)} \phi\ ds + \iint_{(\Sigma)} \psi\ ds.
\label{density functional3}
\end{equation}
In case of equilibrium,   functional  (\ref{density functional3})
must be stationary and  yields   equation of equilibrium  and
 boundary conditions  \cite{Gouin1,Pismen,seppecher}.

\subsection{Equation of equilibrium}
 The
equation of equilibrium is obtained by using the virtual work
principle \cite{germain}. We denote by $\delta\boldsymbol{x}$ the
variation of Euler position $\boldsymbol{x}$ as defined by Serrin
in \cite{serrin}. For
 $\delta \textbf{x}$ null on the  boundaries  of
$L$, the integrals on  $S$ and $\Sigma$ have  null contributions;
the virtual work principle   yields:
\begin{equation*}
\delta \left(\int_{L}\rho\, (\varepsilon +\Omega )\, dv\right)  =
0.
\end{equation*}%
Taking  into account the relation expressing the variation of a
derivative,
\begin{equation*}
\delta \left(\frac{\partial \rho }{\partial \boldsymbol{x}}\right)=\frac{\partial \delta \rho }{%
\partial \boldsymbol{x}}-\frac{\partial \rho }{\partial \boldsymbol{x}}\frac{\partial \delta \boldsymbol{x}}{%
\partial \boldsymbol{x}} ,
\end{equation*}%
with ${\partial }/{\partial \boldsymbol{x}} = {\rm grad}^T $,
where $^T$ denotes the transposition; we obtain:
\begin{equation*}
\delta \beta =2\,\left(\frac{\partial \delta \rho }{\partial
\boldsymbol{x}}-\frac{\partial \rho }{\partial
\boldsymbol{x}}\frac{\partial \delta \boldsymbol{x}}{\partial
\boldsymbol{x}}\right){\rm grad}\,\rho
\end{equation*}%
and%
\begin{equation*}
\delta \varepsilon =\varepsilon '_{\rho }\,\delta \rho
+\varepsilon '_{\beta } \,\delta \beta ,
\end{equation*}
as well as eqs. (14.5) and (14.6) in \cite{serrin}, then
\begin{equation*}
\int_{L}( -\,{\rm div}^T\,\boldsymbol{\sigma} +\rho\,{ \rm grad}\,
\Omega )\, \delta \boldsymbol{x}\, dv=0,
\end{equation*}%
where $\boldsymbol{\ \sigma =}-p\,\boldsymbol{1}-\lambda \;{\rm
grad\ }\rho \ \otimes \ {\rm grad }\rho\,$  is the generalization
of the stress tensor with   $p=\rho ^{2}\varepsilon _{\rho }^{\prime }-\rho \text{ div  }%
(\lambda\,  {\rm grad }\rho )$ \cite{gouin3,trusk}.
 Classical methods
of the calculus of variations   lead to
 the equation of equilibrium:
\begin{equation*}
 \text{div}^T\,\boldsymbol{\sigma} -\rho \  {\rm
grad }\,\Omega   = 0 .\label{motion0}
\end{equation*}
Let us consider   an isothermal vertical film of liquid bounded
respectively by   flat solid wall and   vapour bulk;
$\boldsymbol{i}$ is the upward direction of coordinate
 $x$; then,  gravitational potential   is $\Omega =
 \mathcal{G} \,x$\,, and
\begin{equation}
\text{div}^T\,\boldsymbol{\sigma} - \rho\, \mathcal{G}\,
\boldsymbol{i}=0\, \label{equilibrium1a}
\end{equation}
Coordinate  $z$ is  normal to the  solid wall. Density derivatives
are negligible  in   directions other than  $z$. In liquid-vapour
layer (we also call
 interlayer),
\begin{equation*}
\boldsymbol{\sigma}= \left[
\begin{array}{ccc}
a_1, & 0, & 0 \\
0, & a_2, & 0 \\
0, & 0, & a_3
\end{array}
\right],\quad {\rm with} \quad \left\{
\begin{array}{ccc}
a_1=a_2 = \displaystyle -P+\frac{\lambda }{2}\left(\frac{d\rho
}{dz}\right)^2+\lambda\, \rho\,
\frac{d^2\rho}{dz^2} \\
a_3= \displaystyle -P-\frac{\lambda }{2}\left(\frac{d\rho
}{dz}\right)^2+\lambda\, \rho\, \frac{d^2\rho}{dz^2}
\end{array}\right.
\end{equation*}
Equation (\ref{equilibrium1a}) yields a constant
eigenvalue $a_3$,
\begin{equation*}
P+\frac{\lambda }{2}\left(\frac{d\rho }{dz}\right)^2-\lambda\,
\rho\, \frac{d^2\rho}{dz^2}=P_{v_{b_x}},   \label{equilibrium1b}
\end{equation*}
where $P_{v_{b_x}}\equiv P(\rho_{v_{b_x}})$ denotes the pressure
in   vapour bulk ${v_{b_x}}$ bounding the liquid layer at level
$x$. In the interlayer, eigenvalues $a_1, a_2$ depend on distance
$z$ to the solid wall. In all parts of the isothermal fluid, Eq.
(\ref{equilibrium1a}) can be written \cite{gouin3}:
\begin{equation}
 {\rm grad }\left( \,\mu   -\lambda\, \Delta
\rho + \mathcal{G}\,x\, \right) =0 \,, \label{equilibrium2a}
\end{equation}
where $\mu $ is the chemical potential which is defined to an
unknown additive constant.  We note that Eqs
\eqref{equilibrium1a} and \eqref{equilibrium2a}  are independent
of surface energies (\ref{surfaceenergy}) and (\ref{cl2}). \\ The
chemical potential is a function of $P$  and temperature $T$; due
to  the equation of state, the chemical potential can be also
expressed as a function of $\rho$  and $T$.  We choose as
reference chemical potential $\mu _{o}=\mu _{o}(\rho)$ which is
null for bulks of densities $\rho _{l}$ and $\rho _{v}$ associated
with the phase equilibrium in normal conditions (temperature $T_o$
and atmospheric pressure $P_o$  \cite{Callen}). Due to Maxwell's
 rule,
the volume free energy associated with $\mu _{o}$ is
$g_{o}(\rho)-P_{o}$  where   $ P_{o}\equiv P(\rho _{l})=$ $P(\rho
_{v})$ is the bulk pressure  and $g_{o}(\rho)=\displaystyle
\int_{\rho _{v}}^{\rho }\mu _{o}(\rho)\,d\rho\ $ is   null for the
liquid and vapour  bulks of the phase equilibrium. Pressure $P$ is
\cite{Rowlinson}:
\begin{equation}
P(\rho)=\rho \, \mu _{o}(\rho)-g_{o}(\rho)\ +P_{o}
.\label{therm.pressure}
\end{equation}
Thanks to Eq. (\ref{equilibrium2a}), we obtain in all the fluid:
\begin{equation*}
\mu _{o}(\rho)-\lambda \Delta \rho +   \mathcal{G}\ x =\mu
_{{o}}(\rho _{b}) ,\label{equilibrium2b}
\end{equation*}
where $\mu _{{o}}(\rho _{b})$ is the chemical potential value of
mother liquid-bulk  of density  $\rho _{b}$
 such that $\mu _{{o}}(\rho _{b})= \mu
_{{o}}(\rho_{v_{b}})$; $\rho_{b}$ and $\rho_{v_{b}}$ are the
densities of the
 mother  bulks bounding the layer  at level  $x =0$.
 We
 emphasize that  $P(\rho _{b})$ and $P(\rho_{v_{b}})$ are
unequal as  drop or bubble  bulk pressures. Likewise, we define a
mother liquid-bulk  of density $\rho_{{b_x}}$
 at level  $x$ such that $\mu _{{o}}(\rho _{b_x})= \mu
_{{o}}(\rho_{v_{b_x}})$ with $P(\rho _{b_x}) \neq
 P(\rho_{v_{b_x}})$. Then,
 \begin{equation}
\lambda\, \Delta \rho  =\mu_o(\rho)-\mu_o(\rho _{b_x}) \ \ {\rm
with}\ \ \mu_o(\rho _{b_x}) = \mu_o(\rho _{b}) - \mathcal{G}\,x
\label{equilibrium2c}
\end{equation}
and density derivatives being negligible  in directions other than
$z$, in the interlayer,
 \begin{equation}
\lambda\,\frac{d^2\rho}{dz^2} = \mu_{b_x}(\rho), \quad {\rm
with}\quad \mu_{b_x}(\rho) = \mu_o(\rho)-\mu_o(\rho _{b_x})
\label{equilibrium2d}
\end{equation}
\subsection{Boundary conditions}
Condition  at  solid wall $(S)$   associated with   free
surface energy  (\ref{surfaceenergy}) yields \cite{Gouin1}
\begin{equation}
\lambda \left(\frac{d\rho }{dn}\right)_{|_S}+\phi ^{\prime
}(\rho)_{|_S}\ =0, \label{cl1}
\end{equation}
where $n$ is the external normal direction to the fluid; then,
Eq. (\ref{cl1}) yields
\begin{equation*}
\lambda \left(\frac{d\rho }{dz}\right)_{|_{z=0}}=-\gamma
_{1}+\gamma _{2\ }\rho_{|_{z=0}} . \label{BC1}
\end{equation*}
The condition at   liquid-vapour interface  $(\Sigma)$ associated
with the free surface energy (\ref{cl2}) yields
\begin{equation}
\lambda \left(\frac{d\rho }{dz}\right)_{|_{z=h}}=-\gamma _{4}\
\rho_{|_{z=h}}\,.  \label{BC2}
\end{equation}
In Eq. \eqref{BC2}, density derivative $\,   {d\rho }/{dz}\,$ is
large with respect to the variations of the density in the
interlayer and corresponds to the drop of density in the
liquid-vapour interface. Consequently, Eq. \eqref{BC2} defines the
film thickness   inside the liquid-vapour interface bordering the
liquid layer  at surface $z=h$ considered as a dividing-like
surface  (\cite{Rowlinson}, chapter 3).
\subsection{Disjoining pressure of vertical liquid thin-films}
Equation (\ref{disjoiningpressure}) can be extended to the
disjoining pressure at   level $x$; we obtain the disjoining
pressure value:
\begin{equation*}
\Pi =P_{v_{b_x}}-P_{b_x}\,,  \label{disjoining pressure}
\end{equation*}
where $P_{b_x}$ and $P_{v_{b_x}}$ are the pressures in
mother-liquid and mother-vapour bulks  corresponding to   level
$x$. At a given temperature $T$,   $\Pi$ is a function of
$\rho_{b_x}$ or equivalently a function of $x$. Let us denote by
\begin{equation}
g_{b_x}(\rho) =
g_o(\rho)-g_o(\rho_{b_x})-\mu_{o}(\rho_{b_x})(\rho-\rho_{b_x}),\label{g}
\end{equation}
the primitive of  $\mu _{b_x}(\rho)$ null for $\rho _{b_x}$.
Consequently,  Eq. (\ref{therm.pressure}) gives
\begin{equation}
\Pi (\rho _{b_x}) = -g_{b_x}(\rho_{v_{b_x}}) , \label{disjoining}
\end{equation}
and integration of Eq. (\ref{equilibrium2d}) yields
\begin{equation}
\frac{\lambda }{2}\,\left(\frac{d\rho
}{dz}\right)^2=g_{b_x}(\rho)+\Pi (\rho _{b_x}),
\label{equilibrium2e}
\end{equation}
where $d\rho/dz = 0$ when $\rho = \rho_{v_{b_x}}$.\newline The
reference chemical potential linearized near density $\rho_l$ is
$\ \mu _{o}(\rho)=\displaystyle ({c_{l}^{2}}/{\rho _{l}})(\rho
-\rho_{l})\ $ where velocity $c_l$ is the isothermal
sound-velocity in liquid bulk of density $\rho_l$ at temperature
$T_o$ \cite{espanet}. In the liquid part of the liquid-vapour
film,
 Eq. (\ref{equilibrium2d})   yields:
\begin{equation}
\lambda \frac{d^{2}\rho }{dz^{2}} =\frac{c_{l}^{2}}{\rho
_{l}}(\rho -\rho _{b})+ \mathcal{G}\,x \equiv
\frac{c_{l}^{2}}{\rho _{l}}(\rho -\rho _{b_x})\quad  {\rm with}\ \
\rho _{{b_x}} = \rho _{{b}}-  \frac{\rho _{l}}{c_l^2}\,
\mathcal{G}\, x .\label{liquidensity}
\end{equation}
The reference chemical potential linearized near density $\rho_v$
is $\ \mu _{o}(\rho)=\displaystyle  \frac{c_{v}^{2}}{\rho
_{v}}(\rho -\rho_{v})\ $ where velocity $c_v$ is the isothermal
sound-velocity in vapour bulk of density $\rho_v$ at temperature
$T_o$ \cite{espanet}. In the vapour part    of the liquid-vapour
film, Eq. (\ref{equilibrium2d})   yields:
\begin{equation*}
\lambda \frac{d^{2}\rho }{dz^{2}}=\frac{c_{v}^{2}}{\rho _{v}}(\rho
-\rho _{v_{b}})+\mathcal{G}\,x\equiv \frac{c_{v}^{2}}{\rho
_{v}}(\rho -\rho _{v_{b_x}})\quad {\rm with}\quad  \rho _{v_{b_x}}
= \rho _{v_{b}}- \frac{\rho_v}{c_v^2}\,\mathcal{G}\,x .
\end{equation*}
Due to Eq. (\ref{equilibrium2c}), $\mu _{o}(\rho)$ has the  same
value for $\rho _{v_{b_x}}$ and $\rho _{{b_x}}$; then
\begin{equation*}
\frac{c_{l}^{2}}{\rho _{l}}(\rho _{b_x}-\rho _{l}) =\mu _{o}(\rho
_{b_x})=\mu _{o}(\rho _{v_{b_x}}) =\frac{c_{v}^{2}}{\rho
_{v}}(\rho _{v_{b_x}}-\rho _{v}),   \label{densities1}
\end{equation*}
\begin{equation*}
 {\rm and}\qquad \rho _{v_{b_x}}=\rho _{v}\left(
1+\frac{c_{l}^{2}}{c_{v}^{2}}\frac{(\rho _{b_x}-\rho _{l})}{\rho
_{l}}\right) .  \label{densities2}
\end{equation*}
In   liquid and vapour parts of the interlayer we have,
respectively
\begin{equation*}
g_{o}(\rho)=\frac{c_{l}^{2}}{2\rho _{l}}(\rho -\rho _{l})^{2}\ \ \
{\rm ( liquid)}\quad {\rm and} \quad
g_{o}(\rho)=\frac{c_{v}^{2}}{2\rho _{v}}(\rho -\rho _{v})^{2}\ \ \
{\rm ( vapour)}.
\end{equation*}
From Eqs (\ref{g})-(\ref{disjoining})  we deduce  the disjoining
pressure at level $x$:
\begin{equation}
\Pi (\rho _{b_x})=\frac{c_{l}^{2}}{ 2\rho _{l}}(\rho _{l}-\rho _{b_x})%
\left[ \rho _{l}+\rho _{b_x}-\rho _{v}\left( 2+\frac{c_{l}^{2}}{c_{v}^{2}}%
\frac{(\rho _{b_x}-\rho _{l})}{\rho _{l}}\right) \right] .
\label{disjoining pressure2}
\end{equation}
Due to\ $\ \displaystyle\rho _{v}\left( 2+\frac{%
c_{l}^{2}}{c_{v}^{2}}\frac{(\rho _{b_x}-\rho _{l})}{\rho
_{l}}\right) \ll \rho _{l}+\rho _{b_x}$, we get
$$\displaystyle \Pi
(\rho _{b_x})\approx\frac{c_{l}^{2}}{2\rho _{l}}(\rho
_{l}^{2}-\rho _{b_x}^{2}) . $$
 At level $x=0$,   the  mother liquid-bulk density is closely equal to  $\rho_l$  (the density of liquid in
phase equilibrium). Due to  Eq. (\ref{liquidensity}), $\Pi$ can be
considered as a function of $x$:
\begin{equation}
\Pi
\{x\}\approx\rho_l\,\mathcal{G}\,x\left(1-\frac{\mathcal{G}\,x}{2\,c_l^2}\right).
\label{disjoining pressuregravity}
\end{equation}
We denote $h_x$ in place of $h$ for a vertical film, and we
consider a film   of thickness  $h_x$ at level $x$; the density
profile in the liquid part of the liquid-vapour film is solution
of system:
\begin{equation}
\left\{
\begin{array}{c}
\ \displaystyle\lambda \frac{d^{2}\rho
}{dz^{2}}=\frac{c_{l}^{2}}{\rho _{l}}
(\rho -\rho _{b_x}),  \qquad\qquad {\rm with\ boundary\ conditions}: \\
 \\
\quad \displaystyle \lambda \frac{d\rho }{dz}_{\left|
_{z=0}\right. }=-\gamma _{1}+\gamma _{2\ }\rho _{\left|
_{z=0}\right. }\quad {\rm and}\quad \displaystyle\lambda
\frac{d\rho }{dz}_{\left| _{z=h_x}\right. }=-\gamma _{4}\ \rho
_{\left| _{z=h_x}\right. }.
\end{array}
\right.  \label{systeme1}
\end{equation}
Quantities $\tau $ and $%
d $ are
 \begin{equation*}
 \tau \equiv
\frac{1}{d}=\frac{c_{l}}{\sqrt{\lambda \rho _{l}}}\ , \label{tau}
\end{equation*}
where $d $ is a  reference length;   we introduce   coefficient $
\gamma _{3}\equiv\lambda\, \tau$. The solution of System
(\ref{systeme1}) is
\begin{equation}
\rho =\rho _{b_x}+\rho _{1_x}\,e^{-\tau z}+\rho _{2_x}\,e^{\tau z},
\label{profil}
\end{equation}
where the boundary conditions at $z=0$ and $h_x$ yield the values
of $\rho _{1_x}$ and $\rho _{2_x}$:
\begin{equation*}
\left\{
\begin{array}{c}
(\gamma _{2}+\gamma _{3})\rho _{1_x}+(\gamma _{2}-\gamma _{3})\rho
_{2_x}=\gamma
_{1}-\gamma _{2}\rho _{b_x}, \\
\\
 \quad -e^{-h_x\tau }(\gamma _{3}-\gamma _{4})\rho
_{1_x}+e^{h_x\tau }(\gamma _{3}+\gamma _{4})\rho _{2_x}=-\gamma _{4}\rho
_{b_x}.
\end{array}
\right. \
\end{equation*}
The liquid density profile is a consequence of solution
({\ref{profil}) when   $z$ $\in \left[ 0,h_x\right]$.  By taking
Eq. ({\ref{profil}) into account  in Eq. (\ref {equilibrium2e})
and $g_{b_x}(\rho)$ in  linearized form in the liquid part of the
interlayer, we get
\begin{equation*}
\Pi (\rho _{b_x})=-\frac{2\,c_{l}^{2}}{\rho _{l}}\,\rho _{1_x}\,
\rho _{2_x},
\end{equation*}
 and   consequently,
\begin{eqnarray}
\Pi (\rho _{b_x}) &=&\frac{2\,c_{l}^{2}}{\rho _{l}}\left[ (\gamma
_{1}-\gamma _{2}\rho _{b_x})(\gamma _{3}+\gamma _{4})e^{h_x\tau
}+(\gamma
_{2}-\gamma _{3})\gamma _{4}\rho _{b_x}\right] \times  \notag \\
&&\frac{\left[ (\gamma _{2}+\gamma _{3})\gamma _{4}\rho
_{b_x}-(\gamma
_{1}-\gamma _{2}\rho _{b_x})(\gamma _{3}-\gamma _{4})e^{-h_x\tau }\right] }{%
\left[ (\gamma _{2}+\gamma _{3})(\gamma _{3}+\gamma _{4})e^{h_x\tau
}+(\gamma _{3}-\gamma _{4})(\gamma _{2}-\gamma _{3})e^{-h_x\tau
}\right] ^{2}} .\label{Derjaguine}
\end{eqnarray}
By identification of expressions  (\ref{disjoining pressure2}) and
(\ref {Derjaguine}), we get a relation between $h_x$ and $\rho
_{b_x}$ and  a relation between  disjoining pressure $\Pi (\rho
_{b_x})$ and  thickness  $h_x$ of the liquid film. For the sake of
simplicity, we  again finally denote   the disjoining pressure by
 $\Pi (h_x)$
which is a function of $h_x$ at temperature $T_o$.   \newline
Only Eq. \eqref{disjoining pressuregravity} depends on
$\mathcal{G}$. Due to the fact that  Eq. \eqref{Derjaguine bis}
does not depend on $\mathcal{G}$, its  expression remains
unchanged when we consider $h$ instead of $h_x$.
\newline
Due to  $\rho _{b_x}\simeq \rho _{b}\simeq \rho _{l}$
\cite{Derjaguin}, the disjoining pressure reduces to the
simplified expression
\begin{eqnarray}
\Pi (h_x) &=&\frac{2\,c_{l}^{2}}{\rho _{l}}\left[ (\gamma
_{1}-\gamma _{2}\rho _{l})(\gamma _{3}+\gamma _{4})e^{h_x\tau
}+(\gamma _{2}-\gamma
_{3})\gamma _{4}\rho _{l}\right] \times  \notag \\
&&\frac{\left[ (\gamma _{2}+\gamma _{3})\gamma _{4}\rho
_{l}-(\gamma
_{1}-\gamma _{2}\rho _{l})(\gamma _{3}-\gamma _{4})e^{-h_x\tau }\right] }{%
\left[ (\gamma _{2}+\gamma _{3})(\gamma _{3}+\gamma
_{4})e^{h_x\tau }+(\gamma_{3}-\gamma _{4})(\gamma _{2}-\gamma
_{3})e^{-h_x\tau }\right] ^{2}} \label{Derjaguine bis}\\ \notag
\\
 {\rm with}    \ && \Pi (h_x) \equiv \Pi (h) . \notag
\end{eqnarray}
Let us notice an important property   of any  fluid mixture
  consisting of liquid-water,   vapour-water
and  air \cite{gouin4}. The  mixture's total-pressure is the sum
of  the partial pressures of its components, and at equilibrium
the partial pressure of air is constant through liquid-air and
vapour-air domains. Consequently, results of Section 2 remains
unchanged: the disjoining pressure of the mixture is the same as
for   fluid without  air when only a liquid thin-film  separates
liquid and vapour bulks \cite{Gouin2}.

\subsection{Numerical calculations}

\emph{Mathematica}$^
 {\tiny TM}$ allows us to draw
 the  graphs of  $\Pi(h)$ defined by Eq. \eqref{Derjaguine bis} and
$G(h)$  defined by Eq. \eqref{Gibbs2}   when $h \in
[({1}/{2})\,\sigma_l,\ell]$, where $\ell$ is a few tens
 \AA ngstroms length.
   For a few nanometers, the
 film thickness is not exactly $h$; we must add an estimated thickness
  $2\,\sigma_l$ of   liquid part of the
liquid-vapour interface bordering the liquid layer  and the
 layer thickness is approximatively $ h+ 2\,
\sigma_l$ \cite{Rowlinson}.\\
We considered water at $T_o = 20 {%
{}^\circ}$ C.
\begin{figure}[h]
\begin{center}
\includegraphics[width=7cm]{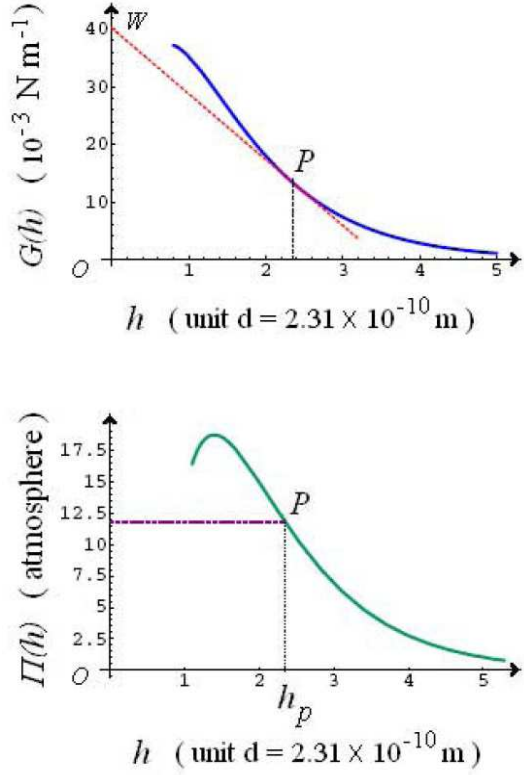}
\end{center}
\caption{{\footnotesize Upper graph:  $G(h)$-graph. The unit of $x-$axis graduated by $h$  is $%
d=2.31\times 10^{-10}$ \texttt{m} ; the unit of $y-$axis is $10^{-3}$ \texttt{N\,m}$^{-1}$ (surface tension).
 Lower graph:
$\Pi(h)$-graph. The unit of $x-$axis graduated by $h$ is
$d=2.31\times 10^{-10}$ \texttt{m}; the unit of $y-$axis is one
atmosphere.}} \label{fig5}
\end{figure}
In S.I. units,   experimental values  are
 \cite{gouin5,Israel,Handbook}:\\
$\rho_l = 998\,  \texttt{kg.\,m} ^{-3}$,  $c_l = 1.478\times
10^{3}\,  \texttt{m.\,s} ^{-1}$,   $c_{ll}=1.4\times 10^{-77}\,
\texttt{kg.\,m}^8.\,\texttt{s}^{-2}$ , $
\sigma _{l}=2.8\times 10^{-10}\,  \texttt{m}$, $%
m_{l}=2.99\times 10^{-26}\, \texttt{kg}$.\newline Silica is
deposited in many plant tissues  such as in   bark and wood. We
choose $\sigma _{s}=2.7\times 10^{-10}\,  \texttt{m}$\,; this
value is intermediate between molecular diameter of silicon and
diameter of non-spherical molecules of water.
\newline
We deduce $\lambda = 1.17 \times 10^{-16}\,
\texttt{kg}^{-1}.\,\texttt{m}^7.\,\texttt{s}^{-2}$,    $d = 2.31
\times 10^{-10}\, \texttt{m}$, $\gamma _2= 5.42 \times 10^{-8}\,
\texttt{kg}^{-1}.\,\texttt{m}^6.\,\texttt{s}^{-2}$, $\gamma _3 =
5.06 \times 10^{-7}\,
\texttt{kg}^{-1}.\,\texttt{m}^6.\,\texttt{s}^{-2}$. \newline
 The
superficial tension of water is $\gamma = 72.5 \times 10^{-3}\,
\texttt{kg.s}^{-2}$.  We choose as Young contact-angle $\theta$
between    xylem wall   and the liquid-water/vapour interface, the
arithmetic average of different Young angles propounded in the
literature; this value is $\theta = 50 $ degrees \cite{Mattia}.
Consequently,
 $\gamma_1$ is deduced from
 the solid-liquid  surface
energy expressed as
$
\phi(\rho_{s})=-\gamma
_{1}\,\rho_{s}+ ({1}/{2})\,\gamma_{2}\,\rho_{s}^{2}.
$ Here $\rho_{s}\simeq \rho_l$ denotes the fluid density value at
the surface. The vapour density is negligible with respect to the water-liquid density and   Young's relation
 \cite{Rowlinson}  yields\
 $
 \gamma \cos\,\theta =  \gamma
_{1}\,\rho_{l}- ({1}/{2})\,\gamma_{2}\,\rho_{l}^{2}
 $ and
we  get
 $\gamma_1 \approx 75 \times 10^{-6}\,\texttt{m}^3.\,\texttt{s}^{-2}$.
 \newline
In the upper graph of Fig.  \ref{fig5}, we present the free energy
graph  $G(h_x)$. Due to $h_x>({1}/{2})\,\sigma_l$, it is not
numerically possible to obtain  the limit point  $W$ corresponding
to the dry wall; consequently, point $W$ is obtained by
interpolation associated with the concave part of the
\emph{G}-curve. To obtain
 pancake thickness $h_p$ corresponding to the smallest thickness of
the liquid layer, we draw  point  $P$,  contact-point of the
tangent line issued from  $W$   to   $G$-curve.
 In the lower graph of Fig. \ref{fig5},   we present the disjoining pressure graph $%
\Pi(h_x)$. The physical part of the disjoining pressure graph corresponding to $%
\partial\Pi/\partial h_x <0$  is associated
with  a liquid layer of several molecules thickness. The non-physical part corresponding to $%
\partial\Pi/\partial h_x >0$ is also  obtained by Derjaguin \emph{et al} \cite{Derjaguin}\,.
The reference length  $d$ is of the same order as $\sigma_l$   and
is a good length unit for very thin-films. The total pancake
thickness  is of one nanometer order corresponding to a
good thickness value for a high-energy surface
\cite{deGennes,Israel}\,; consequently in the tall trees, at high
level, the thickness of the layer is of a few nanometers. The
point $P$  of  the lower graph corresponds to the point $P$ of upper
graph.

\section{Dynamics of liquid thin-films along vertical walls}

The dynamics of liquid thin-films is studied in the isothermal
case.    When $h\ll L$, where $L$ is the
characteristic length along the wall \cite{Gouin7,oron},\\
\emph{i)} The velocity component along the wall is large with
respect to normal
velocity components which can be neglected,\\
\emph{ii)} The velocity value
varies   orthogonally to the wall and it is possible
to neglect velocity spatial derivatives  along the wall   with respect to  normal
velocity derivatives,\\
\emph{iii)} The pressure is constant in
the direction normal to the wall. It is possible to neglect  inertial
term when $Re\ll L/h$, where $Re$ is the Reynolds number of the flow.

The fluid is heterogeneous and the liquid stress tensor is not
scalar. However, it is possible to adapt the results obtained for
viscous flows to motions in liquid thin-films: due to $\ \epsilon
=h/L\ll 1$, we are  in the case of long wave approximation.
\newline We denote the velocity by $\boldsymbol{v}=(u,v,w)$ where
$(u,v)$ are the tangential components to the wall. Due to the fact
that $\,e=\rm{sup}\left( \left\vert w/u\right\vert ,\left\vert
w/v\right\vert \right) \ll 1$, we   are
 in the case of  lubrication approximation. The main parts of
terms associated with second derivatives of liquid velocity
components
correspond to ${\partial ^{2}u}/{\partial z^{2}}$ and ${\partial ^{2}v}/{%
\partial z^{2}}$.\newline
 The density is constant along each stream line ($\overset{%
\boldsymbol{\centerdot }}{\rho }=0\Longleftrightarrow
div\,\boldsymbol{v}=0$) and isodensity surfaces contain the
trajectories. Then, ${\partial u}/{\partial x},{\partial
v}/{\partial y}$ and ${\partial w}/{\partial z}$ have the same
order of magnitude and $\epsilon \sim e$.\newline As in
\cite{Rocard}, we assume that the kinematic viscosity coefficient
$\nu =\kappa/\rho $, where $\kappa$ is the dynamic viscosity, only
depends on the temperature. In motion equation, the viscosity term
is like in liquid bulk \cite{Charlaix}
$$\ ({1}/{\rho })\,\text{div } \boldsymbol{\sigma }_{v} = 2\nu \,%
\left[ \,  \text{div }\boldsymbol{D}\,+ \textbf{\emph{D}}\, {\rm grad \{\thinspace Ln}\,(2\,\kappa)\}\, %
\right],  $$ where $\boldsymbol{\sigma }_{v}$ is the viscous
stress tensor, $\boldsymbol{D}$ is the velocity  deformation
tensor and $\boldsymbol{D}$ grad\{Ln ($2\,\kappa$)\} is negligible
with respect to div$\,\boldsymbol{D}$. \newline In   lubrication
and long wave approximations, the liquid nanolayer motion verifies
\cite{Gouin7,oron}:
\begin{equation*}
{\boldsymbol{a}}+ {\rm grad}[\,\mu _{o}(\rho )-\lambda \,\Delta \rho
\,]=\nu \,\Delta {\boldsymbol{v}}- \mathcal{G}\,{\boldsymbol{i}}\ \ \rm{with}\ \ \Delta {%
\boldsymbol{v}}\simeq
\begin{bmatrix}
\displaystyle\;\frac{\partial ^{2}u}{\partial z^{2}},\displaystyle\;\frac{%
\partial ^{2}v}{\partial z^{2}},0%
\end{bmatrix}%
, \label{motion0}
\end{equation*}%
where $\boldsymbol{a}$ denotes the acceleration vector.
The equation corresponds to  equation of equilibrium ({\ref%
{equilibrium1a}) with additional inertial-term $\boldsymbol a$ and
viscous term $\nu \,\Delta {\boldsymbol{v}}$.\newline In
approximation of lubrication, the inertial term can be neglected
\cite{oron}:
\begin{equation}
 {\rm grad}[\,\mu _{o}(\rho )-\lambda \,\Delta \rho \,]=\nu
\,\Delta {\boldsymbol{v}}- \mathcal{G}\,{\boldsymbol{i}}.
\label{motion}
\end{equation}%
Equation ({%
\ref{motion}) can be separated into tangential and normal components to the solid
wall.

- The normal component  of Eq. (\ref%
{motion}) writes in the same form than for equilibrium:
\begin{equation*}
\frac{\partial }{\partial z}\left[ \ \mu _{o}(\rho )-\lambda \,\Delta \rho \,%
\right] =0,
\end{equation*}%
and consequently,
\begin{equation}
\mu _{o}(\rho )-\lambda \,\Delta \rho =\mu _{o}(\rho^{\ast } _{b_{x}}),
\label{motion h}
\end{equation}%
where $\rho^{\ast } _{b_{x}}$ is the dynamical  mother liquid-bulk
density at level $x$ (different  from $\rho _{b_{x}}$,  mother
liquid-bulk density at level $x $ and at equilibrium, where
quantities    at equilibrium have corresponding
dynamical quantities indicated by $^{\ast }$\,).  \\
A liquid film of thickness $h_{x}^{\ast }$ is associated to
density $\rho^{\ast } _{b_{x}}$. We can write $\mu _{o}(\rho
_{b})-\lambda \,\Delta \rho =\eta (h_{x}^{\ast }),$ where function
$\eta $ is such that $\eta (h_{x}^{\ast })=\mu
_{o}(\rho^{\ast}_{b_{x}})$.

-\ For motions colinear to the solid wall and to the gravity (direction ${\boldsymbol{i%
}}$ and velocity $u\,{\boldsymbol{i}}$), by taking  account of Eq. (\ref%
{motion h})  the tangential component of Eq. (\ref{motion}) writes:
\begin{equation*}
{\boldsymbol{i}}\,.\, {\rm grad{\ }}\mu _{o}(\rho^{\ast } _{b_{x}})=\nu\, \frac{%
\partial ^{2}u}{\partial z^{2}} - \mathcal{G}\,,
\end{equation*}%
which is equivalent to
\begin{equation}
\frac{\partial \mu _{o}(\rho^{\ast } _{b_{x}})}{\partial
\rho^{\ast } _{b_{x}}}\ \frac{\partial \rho^{\ast }
_{b_{x}}}{\partial x}=  \nu\, \frac{\partial ^{2}u}{\partial
z^{2}} - \mathcal{G}.  \label{viscosity}
\end{equation}%
Generally, the kinematic condition at
solid walls is the adherence condition ($u_{z=0}=0$).
Nevertheless, with water flowing on thin   nanolayers \cite{Churaev}, there are
 qualitative observations for slippage  when the Young contact angle is not zero \cite{Blake}. De Gennes said:  \emph{the results led us to think about unusual processes which could take
place near a wall. They are connected with  thickness $h$ of the
film when $h$ is of an order of the mean free path}
\cite{Degennes}. Recent papers in nonequilibrium molecular
dynamics simulations of three dimensional micro Poiseuille flows
in Knudsen regime reconsider   microchannels:  the results point
out that the no-slip condition can be observed for Knudsen flow
when the surface is rough and the surface wetting condition
substantially influences the velocity profiles \cite{Tabeling}. In
fluid/wall slippage, the condition at solid wall writes:
\begin{equation*}
  u=L_{s}\frac{\partial u}{\partial z}  \qquad {\rm at}\  z=0,\
\end{equation*}%
where $L_{s}$ is the  Navier-length \cite{Degennes}. The
Navier-length  may be as large as a few microns \cite{Tabeling}.
At the liquid-vapour interface, we  assume that vapour viscosity
stress is negligible; from continuity of the fluid
tangential-stress through a liquid-vapour interface, we get
\begin{equation*}
 \frac{\partial u}{\partial z}=0 \qquad {\rm at}\ z=h_{x}^{\ast }\;.
\end{equation*}%
Consequently, Eq. (\ref{viscosity}) implies
\begin{equation*}
\nu \,u=\left( \frac{\partial \mu _{o}(\rho^{\ast } _{b_{x}})}{\partial \rho^{\ast }
_{b_{x}}}\ \frac{\partial \rho^{\ast } _{b_{x}}}{\partial x}%
+\mathcal{G}\right) \left( \frac{1}{2}\,z^{2}-h_{x}^{\ast
}\,z-L_{s}h_{x}^{\ast }\right) .
\end{equation*}%
At
level $x$, the mean spatial velocity $\overline{u}$ of the liquid in the nanolayer is
\begin{equation*}
{\displaystyle\overline{u}=\frac{1}{h_{x}^{\ast }}\int_{o}^{h_{x}^{\ast }}u\
dz}
\end{equation*}%
and consequently,
\begin{equation*}
\nu \,{\boldsymbol{\overline{u}}}=-h_{x}^{\ast }\left( \frac{h_{x}^{\ast }}{3}%
+L_{s}\right) \ \left[ \, {\rm grad}\ \mu _{o}(\rho^{\ast } _{b_{x}})+\mathcal{G}\,\textbf{i}%
\right] \qquad \rm{with}\qquad {\boldsymbol{\overline{u}}}=\overline{u}\ {%
\boldsymbol{i}}\,.
\end{equation*}%
Let us note that:
\begin{equation*}
\frac{\partial \mu _{o}(\rho^{\ast } _{b_{x}})}{\partial x}=\frac{\partial
\mu _{o}(\rho^{\ast } _{b_{x}})}{\partial \rho^{\ast } _{b_{x}}}\,\frac{%
\partial \rho^{\ast } _{b_{x}}}{\partial h_{x}^{\ast }}\,\frac{\partial
h_{x}^{\ast }}{\partial x}\equiv \frac{1}{\rho^{\ast } _{b_{x}}}\,\frac{%
\partial P(\rho^{\ast } _{b_{x}})}{\partial \rho^{\ast } _{x}}\,\frac{%
\partial \rho^{\ast } _{b_{x}}}{\partial h_{x}^{\ast }}\,\frac{\partial
h_{x}^{\ast }}{\partial x}.
\end{equation*}%
Due to the fact the vapour-bulk pressure $P^{\ast }_{v_{b_{x}}}$
is   constant along  the xylem tube,   by using relation $\Pi
(h_{x}^{\ast })=P^{\ast }_{v_{b_{x}}}-P^{\ast }_{b_{x}}$, we get
along the flow motion
\begin{equation*}
\frac{\partial \mu _{o}(\rho^{\ast } _{b_{x}})}{\partial x}=-\frac{1}{\rho^{\ast }
_{b_{x}}}\ \frac{\partial \Pi (h_{x}^{\ast })}{\partial h_{x}^{\ast }%
}\ \frac{\partial h_{x}^{\ast }}{\partial x}
\end{equation*}%
and consequently,
\begin{equation}
\chi^{\ast } _{b_{x}}{\boldsymbol{\overline{u}}}=h_{x}^{\ast }\left( \frac{%
h_{x}^{\ast }}{3}+L_{s}\right) \left[ \, {\rm grad}\ \Pi (h_{x}^{\ast })- {\rho^{\ast }
_{b_{x}}}\,\mathcal{G}\,\textbf{i}%
\right] ,  \label{variation potentiel chimique}
\end{equation}%
where $\chi^{\ast } _{b_{x}}=\rho^{\ast } _{b_{x}}\nu\, $ is the
liquid dynamic-viscosity.\newline
 The mean liquid velocity is
driven by   variation of the disjoining pressure (and
 film thickness) along the solid wall. Equation (\ref{variation potentiel chimique}) differs from   classical   hydrodynamics; indeed, for a classical liquid
thin-films, the Darcy law is ${\boldsymbol{\overline{u}}}=-K(h)\,\rm{%
grad}\,\wp$, \ where $\wp$ is the liquid pressure and $K(h)$ is
the permeability coefficient. In Eq. (\ref{variation potentiel
chimique}), the sign is opposite and the liquid pressure is
replaced by the disjoining pressure.  We note that $ \chi^{\ast }
_{b_{x}}\simeq \chi$\,, where $\chi$ is the liquid kinetic
viscosity in the liquid bulk at phase equilibrium \cite{Charlaix}.
Moreover   $h_{x}^{\ast }/L_{s}\ll 1$, and slippage
 is strongly different from   the adherence condition  corresponding to  $L_{s}=0$.\newline
The  averaged mass equation over the liquid depth is
\begin{equation*}
\frac{\partial }{\partial t}\left( \int_{0}^{h_{x}^{\ast }}\rho \,dz\right) +%
\rm{{div}\left( \int_{0}^{h_{x}^{\ast }}\rho \,\boldsymbol{u}\,dz\right) =0.}
\end{equation*}%
Since the variation of density is small in the liquid nanolayer, the
equation for the free surface is
\begin{equation}
\frac{dh_{x}^{\ast }}{dt}+h_{x}^{\ast }\ \rm{div}{\ \boldsymbol{\overline{u}}%
}=0.  \label{h-equation}
\end{equation}%
By replacing (\ref{variation potentiel chimique}) into (\ref{h-equation}) we
 get
\begin{equation}
\frac{\partial h_{x}^{\ast }}{\partial t}+\frac{1}{\chi}\  %
\rm{div}\left\{ h_{x}^{\ast 2}\left( \frac{h_{x}^{\ast }}{3\,}%
+L_{s}\right)  \Big[\,\rm{{grad}\,\Pi (h_{x}^{\ast })-{\rho^{\ast
} _{b_{x}}}\, \mathcal{G}\,\textbf{i}\Big]}\right\} =0,
\label{h-evolution equation}
\end{equation}
where $\rho^{\ast }
_{b_{x}}\simeq \rho_l$.
Equation (\ref{h-evolution equation}) is a non-linear
parabolic equation.  If ${ \displaystyle \partial \Pi (h_{x}^{\ast })/
\partial h_{x}^{\ast }<0}$ the flow is stable. This result is in accordance
with the static criterium of stability  for liquid thin-films. \newline
When $L_{s}\neq 0$, we notice the flow is multiplied by the factor $%
1+3L_{s}/h_{x}^{\ast }$. For example, when $h_{x}^{\ast }=3\,\texttt{nm}$ and $%
L_{s}=100\,\texttt{nm}$ which is a Navier length of small magnitude with respect to
experiments, the multiplier factor is $10^2$; when $L_{s}$ is $7\,\boldsymbol{\mu}\texttt{m}$ as
considered in \cite{Tabeling}, the multiplier factor is $10^{4}$, which seems possible
 in nanotube observations \cite{Gouin9,Mattia}.\\

 \noindent In  following sections, we use previous tools to study
  watering of  plants and especially trees.

\section{The tree watering}

\subsection{Experiments and  analyzes}

 Since the beginning of the cohesion-tension
 theory, many efforts have   been done to understand
 crude sap motions and to replicate
  tree functions when vessels are under tension.  Synthetic systems simulating
   transport processes have played an important role in  model testing,
   and methods creating microfluidic structures to mimic tree vasculature have been developed
     to capture fundamental aspects of flows  and  xylem tension \cite{Stone,Wheeler}.

When xylem tubes are completely filled with sap,   flows  along
vessels can be compared with   flows through capillaries \cite
{Zimm}.
 Adjacent xylem walls  are connected by  active bordered-pit membranes
   with micropores \cite{Meyra}. The  membranes   separate  two volumes of fluid,
   and generally refer   to  lipid bilayers that surround living cells or
intracellular compartments \cite{Stroock}.  The micropores are a
few tens of microns wide. Due to the meniscus curvatures   at
micropore apertures, marking off liquid-water bulk from air-vapour
atmosphere, the water-bulk pressure is negative inside micropore
reservoirs, but, surprisingly,  semi-permeable micropores allow
flows of liquid-water at negative pressure  to be pushed toward
air-vapour domains at positive pressure   \cite{Tyree2,Tyree3}.
Bubbles spontaneously appear from germ existing in crude sap and
cavitation makes some tubes embolized \cite{Canny,Choat}. Due to
experiments described in Section 2, liquid  thin-films must damp
embolized xylem walls; consequently,  thin-films and microtubes
filled of crude sap are in competition.

Optical measurements indicate  Young's contact-angles of about 50${%
{}^\circ}$ for water on the xylem at 20${%
{}^\circ}$ Celsius \cite{Mattia}. This value suggests that   xylem
walls  are not fully wetting and the capillary spreading cannot
really aid the liquid-water refilling but may explain the apparent
segregation of liquid-water into droplets   \cite{Zwieniecki2}.
The crude sap is not pure water; its liquid-vapour surface tension
has a lower value than the surface tension of pure water   and it
is possible to obtain the same
spreading coefficients with less energetic surfaces.\\
Water exits  the leaves by evaporation through stomata into
subsaturated air. Resistance of  stomata sits in the path of
vapour diffusion between the interior surfaces of leaves and the
atmosphere
 but many of the tallest trees appear to lack active loading mechanisms
   \cite{Fu}. When active transpiration occurs, stomata are open and these pumps run.
   The growth and degrowth of bubbles are rapid within xylem segments, but at night,
    although the stomata are closed,   xylem vessels  developing embolies during the day
       can be refilled with liquid-water and the metastability of the liquid-water may persist
       even in the absence of transpiration   \cite{Zufferey,Zwieniecki1}. \\
Recent advances in tree hydraulics have demonstrated that,
contrary to what was previously believed, embolism and repair may
be far from routine in trees. Trees can recover partially or
totally from the deleterious effects of water stress until they
reach a lethal threshold of cavitation  \cite{Delzon}.  This
result
 can be related with the fact that thin films with a thickness greater
 than the pancake-layer's one are  stable and the behaviour is different
  from bubble stability, which is associated with a saddle point \cite{Slemrod}.

\subsection{ Motions in filled microtubes  and in   thin-films}
\subsubsection{ Generality}
One   important design requirement  is that vapour blockage does
not happen in the stems. When the vessel elements are tight-filled
with crude sap, liquid motions
  are  Poiseuille flows   \cite{Zimm}.
The flow rate through  capillary tubes is proportional to the
applied pressure gradient, the hydraulic conductivity and  depends
on the fourth power of capillary radius \cite{batchelor}. To be
efficient for  sap transportation, the tubes' diameters should be
as wide as possible; because of the micron  size of the xylem
tubes, this is not the case. Consequently, the  tracheary
elements' network must be important. But the sap movement is
induced by  transpiration across micropores located in tree leaves
and the transpiration is bounded by  micropores' sizes; it seems
natural to surmise that the diameters of vessels must  not be too
large to generate a sufficient sap movement.

When the vessel elements are embolized,  thin-films damp the xylem
walls and Eq. (\ref{variation potentiel chimique}) governs the
liquid motion. The diameters of capillary vessels  which range from 10
to 500 $\boldsymbol{\mu} \texttt{m}$ and liquid thin-films of some
nanometre thickness can be considered as plane interlayers.    It
is noticeable that trees can avoid having very high energy
surfaces: if we replace the flat surfaces of the vessels
  with corrugated surfaces at molecular scale,  it is much easier to obtain the
complete wetting requirement, which is otherwise only partial.
However, they are still   internally wet  if crude sap flows
through wedge-shaped corrugated pores. The wedge does not have to
be perfect on the nanometric scale to significantly enhance the
amount of liquid flowing at modest pressures, the walls    being
 endowed with an average
surface-energy \cite{Finn,Wenzel}.

It is interesting to compare    liquid motions both in
tight-filled vessels and   liquid-water thin-films. An hydraulic
Poiseuille flow is \emph{rigid} due to the liquid
incompressibility, the pressure effects are fully propagated in
the tube. For a thin layer flow, the flow rate can increase or
decrease due to the spatial derivative of $h_{x}^{\ast }$ and
depends on the  local disjoining pressure.
 The tree's  versatility allows it to adapt to the local
 disjoining-pressure gradient effects by opening or closing
  the stomata and the curvature of pit pores, so that the bulk pressure
   can be more or less negative and   the transport of water  can be differently
dispatched through the stem parts.

\subsubsection{Numerical calculations}
The treachery network of xylem microtubes is extremely developped.
 For   tree hight   $H = 20\, \texttt{m}$, the total area of xylem walls can be estimated
 to $S = 30\, \texttt{km}^2 \equiv 3 \times 10^7\, \texttt{m}^2$ \cite{Zimm}. We consider
 xylem microtubes with   diameter $2R = 50\, \boldsymbol{\mu}\texttt{m}
 \equiv 5 \times 10^{-5} \texttt{m}$. The dynamic viscosity of liquid water at 20${%
{}^\circ}$ Celsius is $\chi = 10^{-3} \,\texttt{kg.m}^{-1}.
\texttt{s}^{-1}$. It is experimentally verified that in
tightly-filled microtubes, the crude sap velocity usually goes
from $1\, \texttt{m.hour}^{-1} \equiv 2.8 \times 10^{-4}
\texttt{m.s}^{-1}$ to $100\, \texttt{m.hour}^{-1} \equiv 2.8
\times 10^{-2} \texttt{m.s}^{-1}$  in  maximal transpiration
\cite{Zeppel}. The mean velocity of Poiseuille's flows verifies
\begin{equation*}
{\overline{u}}_1 =  \frac{R^2}{8 \chi} \ \left|{\rm{grad}}\,
\left(-\wp_{b^*_x}\right)\right|
 ,
\end{equation*}
where $\wp_{b^*_x}$ is the liquid pressure at level $x$.
Consequently, $\left|{\rm{grad}}\,
\left(-\wp_{b^*_x}\right)\right|$ goes from $ 3.6 \times 10^{-2}
\texttt{ atmosphere.m}^{-1} \equiv 3.6 \times 10^{3}\,
\texttt{Pa.m}^{-1} $ to $ 3.6 \texttt{ atmosphere.m}^{-1} \equiv
3.6 \times 10^{5}\, \texttt{Pa.m}^{-1} $. The number of xylem
microtubes can be calculated as $N=S/(2\,\pi\,R\, H)$; we
approximatively obtain $N= 10^{10}$ microtubes and the total flow
is
\begin{equation*}
Q_1 =  N \frac{\pi R^4}{8\,\chi} \ \left|{\rm{grad}}\,
\left(-\wp_{b^*_x}\right)\right|.
\end{equation*}
For a velocity ${\overline{u}}_1 = 1\, \texttt{m.hour}^{-1} \equiv
2.8 \times 10^{-4}\, \texttt{m.s}^{-1}$,  $Q_1= 5.5 \times 10^{-3}
\texttt{m}^{3}.\texttt{s}^{-1} \equiv 5.5\  \texttt{liters\,/s}$
and for ${\overline{u}}_1 = 100\  \texttt{m.hour}^{-1} \equiv 2.8
\times 10^{-2}\  \texttt{m.s}^{-1}$,  $Q_1= 5.5 \times 10^{-1}\
\texttt{m}^{3}. \texttt{s}^{-1} \equiv 550\ \texttt{liters/s} $
corresponding to  values of biological experiments.
\\
In embolized microtubes,  mean velocity $u_2$ along thin-films is
given by Rel \eqref{variation potentiel chimique}. Pressure
$P_{v_{b_x}} \approx P_v$ being constant and ${\rho^{\ast }
_{b_{x}}} \approx \rho_l$, then $${\rm{grad}} \left( \Pi
(h_{x}^{\ast })- {\rho^{\ast } _{b_{x}}}\,\mathcal{G}\,
x\right)\approx{\rm{grad}}
\left(-P_{b^*_x}-\rho_l\,\mathcal{G}\,x\right).$$ From
$\chi\,{\overline{u}}_2 \approx h_x L_s \, \left|{\rm{grad}}
\left( \Pi (h_{x}^{\ast })- {\rho _{b^{\ast }_{x}}}\,\mathcal{G}\,
x\right)\right|$, together with a layer thickness $h_x = 10\
\texttt{nm} \equiv 10 ^{-8}\, \texttt{m}$, $L_s = 7 \,
{\boldsymbol\mu}\texttt{m} \equiv 7 \times 10^{-6}\, \texttt{m}$,
and $\left|{\rm{grad}} \left( \Pi (h_{x}^{\ast })- {\rho _{b^{\ast
}_{x}}}\,\mathcal{G}\, x\right)\right| = 3.6 \times 10^5 \
\texttt{Pa.m}^{-1}$, we obtain ${\overline{u}}_2 =
 2.5 \times 10^{-5}\  \texttt{m.s}^{-1}$
which is $11$ times less than velocity in tight-filled microtubes
for a velocity  of 1$\  \texttt{m.hour}^{-1}\, \texttt{m}  \equiv
2.8 \times 10^{-4} \texttt{m.s}^{-1}$.\newline For N microtubes,
the total flow is
\begin{equation*}
Q_2 = N 2\pi R h_x u_2
\end{equation*}
and for $10^{10}$ microtubes, $Q_2 = 4 \times 10^{-7}\,
\texttt{m}^3.\texttt{s}^{-1} \equiv 0.4 \,
\texttt{cm}^3.\texttt{s}^{-1}$ which is very small with respect to
the sum of Poiseuille's flows.
\newline
It seems that the liquid thin-films do not affect   the watering
of trees but in next Section 6,  we see   it is not the case. For
an ascent of $50\ \texttt{m}$, we obtain an ascent time of $ 2
\times 10^6 \, \texttt{s} \equiv 23 $ days which estimates the
tree-recovery time in spring.

\section{Embolization and recovery}

\subsection{A diagram of  vessel elements for tall trees}

 \begin{figure}
\begin{center}
\includegraphics[width=7.22cm]{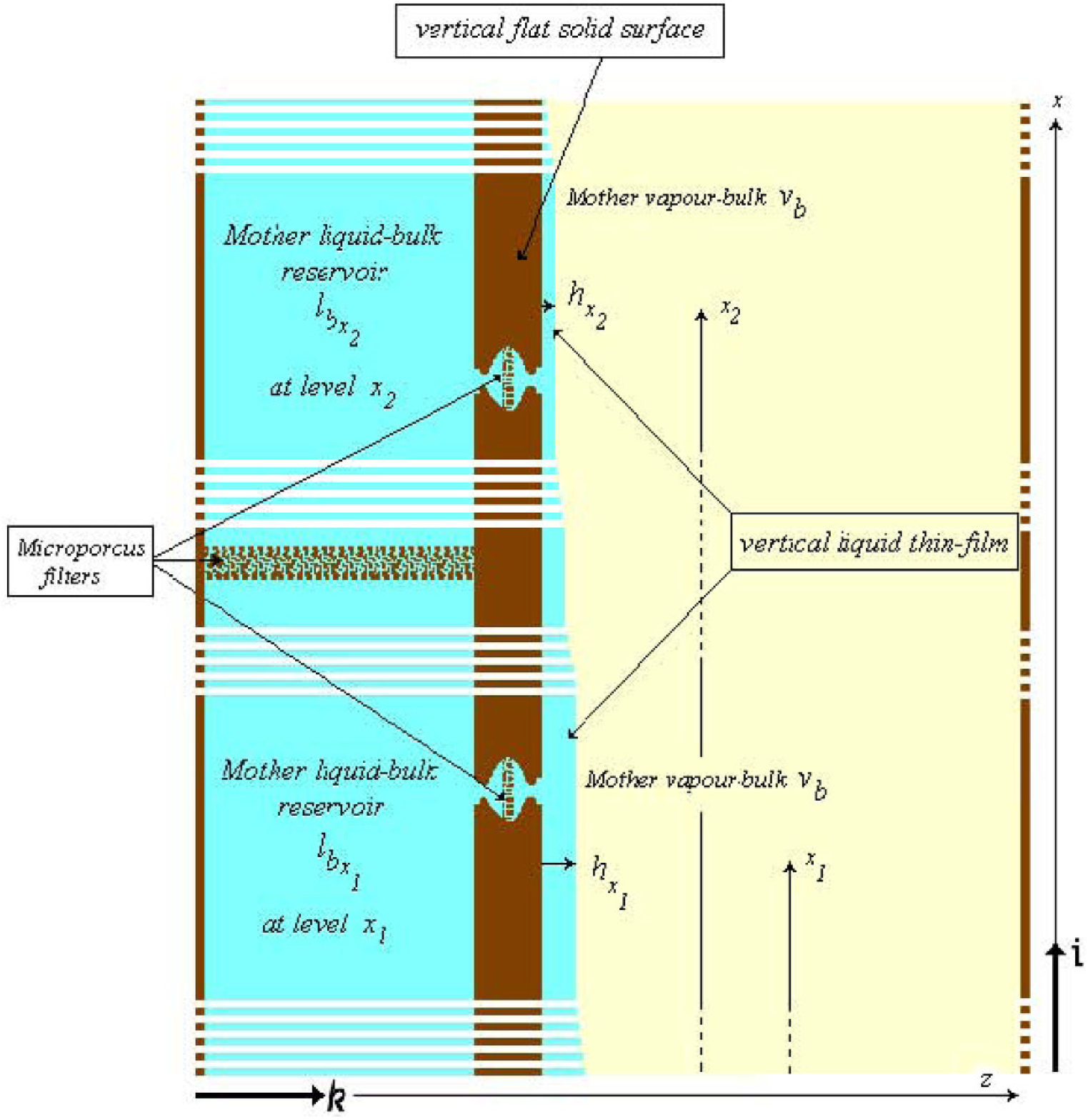}
\end{center}
\caption{\footnotesize{Diagram of a vertical liquid thin-film.  A
liquid thin-film bordered by a mother
 vapour-bulk $(v_b)$ wets
 a vertical flat solid surface. Mother liquid-bulk $(l_{b_x})$, but not mother vapour-bulk $(v_{b})$,
 depends on the x-level.  Like on Sheludko's apparatus,
microporous filters connect the mother liquid-bulk reservoir at
different altitudes with the liquid thin-film. Different parts of
the mother liquid-bulk reservoir can    also be  connected by
microporous filters.}} \label{fig6}
\end{figure}
In  physical conditions of Subsection 3.4 and   temperature at
$20^{\circ}$ Celsius, we consider two vertical adjacent vessels
linked by micropore reservoirs   with pit membranes dotting their
walls. The bordered-pit membranes   are of the order of few tens
of microns corresponding to the microporous filters in Fig.
\ref{fig6}. The mother
 vapour-bulk contains air and the mother liquid-water bulk also  contains dissolved
air.
\\
One  vessel - corresponding to subsaturated mother air-vapour bulk - is embolized
with a  positive pressure;
 it generates a liquid thin-film   which wets the xylem wall.
The other vessel   is filled with the mother liquid-water bulk
at a negative pressure linked to the liquid  thin-film
 thanks to a micropore reservoir with a bordered-pit membrane.

Such a system can be in equilibrium, although the pressure is not the same in the two adjacent vessels.\\
In the same configuration, the  vessel elements are now assumed to be weakly out-of-equilibrium.
 As explained in Section 4, the driving force of
  the sap ascent comes from the decreasing thickness of the liquid thin-film wetting the walls in the embolized vessels;
   consequently the negative pressure value  of the mother liquid-water bulk in
    micropore reservoirs  decreases  (its absolute value increases). Additionally, air-vapour pockets can coexist with the
    mother liquid-water bulk of one of the
    two vessels. The air-vapour pockets  also generate  liquid   thin-films bordering  xylem walls (see Fig. \ref{fig7}).\\
  Due to their curvature,  the pressures of air-vapour pockets are generally higher than the
  mother  vapour-bulk pressure
in the other vessel. The vapour pockets and their liquid
thin-films   empty into the  vessel  with the lower air-vapour
bulk pressure. The analogy proposed in Section 2 between liquid
thin-films and liquid-vapour interfaces of bubbles allows to
simply understand the directions of motions  between air-vapour
pockets and the mother air-vapour bulk: for example, when two
bubbles are included in a liquid bulk (corresponding to the mother
liquid-water bulk at
 negative pressure),
the smallest bubble with the greatest pressure (i.e.
the air-vapour pocket) empties into the largest bubble
with the lowest pressure (i.e. the air-vapour bulk of the embolized vessel) \cite{McCarthy}.

\noindent Such  events  happen  in particular at night, when - due
to the absence of evaporation - the vapour is subsaturated in
embolized vessels;
  the curvature of the air-vapour pockets generates a  pressure greater than the pressure in the embolized vessels.
  Conversely, during the day and strong sunlight, the vapour in  vessels is saturated
  by evaporation; the air-vapour pressure increases in the embolized vessels and the air-vapour gas must
   flow  back into the
  vessels with  air-vapour pockets of subsaturated vapour included in the mother liquid-water bulk at  negative pressure.
\\
It is not surprising that the heartwood may contain liquid under
positive pressure while in the sapwood the transpiration stream
moves along a gradient of negative pressure.  Embolized vessels
creating thin-films may provide a key contribution to  tree
refilling. Crude sap in the heartwood can also fill the vessel
elements at negative pressure through the bordered-pit membranes.
Consequently, embolized microtubes of xylem fundamentally
contribute to the crude-sap ascent and to the refilling of the
tree as machine allowing to obtain   equilibrium between fluid
phases   at different pressures and consequently to recover the
water from cavitation.
\begin{figure}
\begin{center}
\includegraphics[width=9cm]{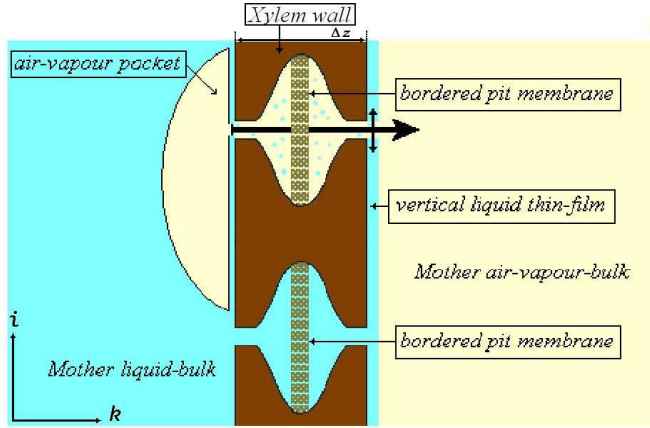}
\end{center}
\caption{\footnotesize At equilibrium between bulks at altitude
$x$, mother liquid-bulk $({l_{b_x}})$ balances the liquid
thin-film and mother air-vapour-bulk $({v_{b}})$. It is not true
for the vapour pocket and the mother air-vapour-bulk: when the
pressure in the air-vapour pocket is greater
 than the pressure in mother air-vapour-bulk, the air-vapour pocket empties
  into the mother air-vapour-bulk.}
\label{fig7}
\end{figure}

\subsection{Numerical values of the mass transfer}

 It is interesting to estimate the magnitude of  air-vapour flows
between air-vapour pockets and the mother air-vapour bulk. At
altitude $x$, the  mother bulks are still named $({l_{b_x}})$ and
$({v_{b}})$, respectively.
 At equilibrium between the two mother
bulks, the vertical  liquid thin-film between  the mother
air-vapour bulk and the xylem wall  is also at equilibrium
 and at the same pressure as  the mother air-vapour bulk. Depending on reservoir altitude $x$,
 the mother liquid-bulk can be at negative pressure.

 The channels crossing from the left   to  the right parts of the xylem wall can
 be considered as  sound pipes  (Fig.\,\ref{fig7}).  We denote by $e$ and
   $2r$ the  length and diameter  of  pipes, respectively.
   The pipe diameters are assumed to be of the same order than
   diameters of  pits bordering the xylem walls. The difference of pressure
    between air-vapour pockets and the mother air-vapour bulk is denoted by $\triangle P$.
    The flow is assumed to be  Poiseuille flow and the mean velocity along the channel is \cite{batchelor}:
\begin{equation}
{\overline{v}} = \frac{r^2}{8 \chi} \,    \frac{\triangle P}{e} .
\label{pulse}
\end{equation}
The velocity is null in  the mother air-vapour bulk and in the
air-vapour pockets creating a velocity-pulse between the two
extremities of the pipe.

On one hand, we assume that the pressure in the air-vapour
  pocket is    greater
  than the  pressure in the mother
   air-vapour bulk. The air-vapour pocket
   and the mother air-vapour bulk are not at equilibrium.
   The liquid thin-film between
   the air-vapour pocket and the xylem wall is thinner than
   the liquid thin-film between  the mother air-vapour bulk and the xylem wall   (see Fig. \ref{fig7}).\\
   We consider experimental values:\newline
   $e = 50\, \boldsymbol{\mu}\texttt{m} \equiv 5  \times
10^{-5}\, \texttt{m}$\,,
    $r = 10\, \boldsymbol{\mu}\texttt{m} \equiv 1 \, \times 10^{-5}\,
     \texttt{m}$\,, $\chi = 1.81 \times 10^{-5} \, \texttt{kg.m}^{-1}.
      \texttt{s}^{-1}$ for air-vapour viscosity  and the saturated vapour pressure of water  is
       $\ 2.3 \times 10^{-2}\, \texttt{atmosphere} \equiv 2.3 \times 10^3\, \texttt{Pa}$ at  20${%
{}^\circ}$ Celsius \cite{Handbook,Zimm}. In the case of a
difference of pressure between the saturated vapour pressure in
air-vapour pocket and the
 sub-saturated vapour pressure in the mother air-vapour bulk
$\triangle P = 10^{-2}\, \texttt{atmosphere} \equiv 10^{3}\, \texttt{Pa}$, we obtain ${\overline{v}}
 = 14\, \texttt{m.s}^{-1}$ and the crossing time of the pipe is $\tau= e/{\overline{v}}
  = 3.6 \times 10^{-6}\, \texttt{s}$ corresponding to a pulse frequency $\omega =\tau^{-1}
  \approx 2.8\,\times 10^{5}\, \texttt{Hz}\equiv 2 80\,  \texttt{KHz}$ associated with ultrasonic vibrations. \\
It is interesting to calcutate  the flow   through the pipe,
\begin{equation*}
q =  \frac{\pi r^4}{8\,\chi} \, \frac{\triangle P}{e}.
\end{equation*}
 We obtain $q = 4.3
\,\times 10^{-9}\, \texttt{m}^3.s^{-1} \equiv 4.3\times 10^{-3}\,
\texttt{cm}^3.s^{-1}$.
\\
  Conversely,
  when the  mother air-vapour pressure is slightly greater than the
  air-vapour pocket pressure, the air-vapour bulk can embolize the liquid tight-filled  vessel elements
  with opposite velocity  and consequently the same pulse frequency for the same difference of pressure
  between mother air-vapour-bulk and air-vapour pockets.\\
  Let us note that the saturated vapour pressure quickly decreases
  with the temperature. At  0${%
{}^\circ}$ Celsius, the saturated vapour pressure of
 water is $\  6 \times 10^{-4} \,\texttt{atmosphere} \equiv 60
    \, Pa  $\ \, and
    $\triangle P$ drastically decreases as the flow through the
    pipe, and it may be a reason of embolization at low
    temperature.

On the other hand,   the crude sap in the tight-filled microtubes
can spread in the embolized microtubes when the disjoining
pressure is smaller than the disjoining pressure at equilibrium.
Equation \eqref{pulse} allows to calculate the mean velocity of
the flow:  here $\triangle P$ represents the difference between
the disjoining pressure values at and out from equilibrium.
\newline With $\triangle P = 10^{-1}\, \texttt{atmosphere} \equiv
10^{4}\, \texttt{Pa}$\, and $\chi = 10^{-3}\, \texttt{kg.m}^{-1}$,
we obtain ${\overline{v}} = 2.5  \ \texttt{m.s}^{-1}$ and the
crossing time of the pipe is $\tau= e/{\overline{v}} = 2 \times
10^{-5}\, \texttt{s}$\ corresponding to a pulse frequency $\omega
=\tau^{-1} \approx 5\,\times 10^{4}\, \texttt{Hz}\equiv 50\,
\texttt{KHz}$. We also obtain  $q = 0.8 \,\times 10^{-9}\,
\texttt{m}^3.s^{-1} \equiv 0.8\, \,\times
10^{-3}\,\texttt{cm}^3.s^{-1}$.

Due to the size of xylem microtubes, a reference diameter of
bubbles can be about $2R = 50\, {\boldsymbol\mu}\texttt{m}\equiv 5
\times 10^{-5}\, \texttt{m}$\ corresponding to   volume $V
=6.5\times 10^{-14}\, \texttt{m}^3\equiv 6.5\times 10^{-8}\,
\texttt{cm}^3$ and for both liquid and gas exchanges,  the
transfers of masses are extremely fast.\newline

 The fast accelerations of air-vapour gas through micropores   generate ultrasounds associated with pulse frequencies and may explain
the acoustical measurements obtained in experiments  \cite{Milburn,Tyran}.\\

  The magnitude of the viscosity of simple wetting
fluids increases  when they are
confined between solid walls,  and there is a direct correlation between the air-seeding threshold and the pit pore membranes' diameters \cite{Jansen}.
However, the acceleration magnitude is so large that it remains very important for
viscous fluids and   semi-permeable bordered-pit membranes.

\subsection{Limit of the disjoining pressure model and topmost
trees} Liquid thin-films  primarily contribute to
 xylem microtubes refilling and consequently the  refilling is
not possible   when the liquid thin-films break  down. Amazingly,
the above study  allows to estimate a maximum of thin-films'
altitude.    In Subsection 3.5 and in Fig. \ref{fig5}, we  have
seen that the thickness of the liquid thin-film decreases when its
altitude increases and
  the liquid thin-film disrupts when thickness  reaches the pancake layer thickness-value.

  The pancake layer of thin-films was presented in Section 2 and
  a numerical simulation associated with experimental data of xylem  at temperature $20 {%
{}^\circ}$ Celsius is presented
  in Subsection 3.5.
  As  indicated in Subsection 3.5,
the Young contact angle  between a xylem wall and a liquid-vapour
water interface is $\theta \approx 50{{}^\circ}$.
\\
The upper graph in Fig.\ \ref{fig5}  presents the free energy
   $G(h)$ associated with physical values of  trees' xylem
walls.  The lower graph of Fig.\,\ref{fig5},
 presents the disjoining pressure   $\Pi(h)$ and
  is in accordance with
experimental curves obtained in the literature
\cite{Derjaguin,Gouin8}. The total pancake thickness $h_p$ is
about one nanometer order corresponding to a good thickness value
for a high-energy surface \cite{deGennes}; consequently in   tall
trees, at high level, the thickness of the liquid thin-film must
be of a few nanometers. Point \emph{P} on the lower graph
corresponds to point \emph{P} on upper graph.
 When $x_{_{P}}$ corresponds to the altitude of the
pancake layer, Eq. \eqref{disjoining pressuregravity} and
$\mathcal{G}x/(2\,c_l^2) \ll 1$ yields $\Pi(h_p) \simeq \,
\rho_l\,\mathcal{G}\,x_{_{P}}$. From the lower graph in Fig.\
\ref{fig5}, we obtain  a maximum thin-film height of  $120$ meters
corresponding to $12$ atmospheres. At this altitude, we must
approximatively add 20 meters corresponding to the ascent of sap
due to capillarity and osmotic pressure \cite{Callen,Ruggeri} and
we obtain 140 meters. This level corresponds to the level order of
the topmost trees, as a giant, 128 meter-tall
 eucalyptus or a 135 meter-tall sequoia which were reported  in the past by Flindt \cite{Flindt}.
\\
Other mechanical or biological constraints may suggest adaptation
to height-induced costs  \cite{Koch}, but nevertheless our model
limits the maximum height of trees. The tallest trees are not the
ones with the largest demand for tension; it is rather dry climate
shrubs that demand it \cite{Benkert}. This observation seems to be
in accordance with the possible existence of thin-films in
embolized vessels at high elevation.

\section{Conclusion}

In the trees, xylem microtubes are naturally filled with sap up to
an altitude of a few ten meters.
  Above this altitude, when  xylem tubes can be embolized, the molecular forces
     create crude sap thin-films along the walls of xylem associated with   micropore pressures
  versatilely adapted thanks to pit membranes.   The
disjoining pressure of liquid thin-films is the exhaust valve
filling the  xylem microtubes and allowing  crude sap to ascend.
Consequently, the embolized vessels constitute a necessary network
for the watering and the recovery of  tall trees (Lampinen and
Noponen  argued that embolisms were necessary for the ascent of
sap \cite{Lampinen}).  The model  explains aspects of sap movement
which the classical cohesion-tension theory was hitherto unable to
satisfactorily account for, e.g. the refilling of the vessels in
spring, in the morning  or after embolism events, as well as the
compatibility with thermodynamics' principles. Nevertheless, the
xylem tight-filled microtubes under tension are the essential
network of tree watering. \newline Simple in vivo observations at
the nanometre thickness of  liquid  thin-films   are not easy to
implement
 and the  direct measurement difficulties  prevent their detection. The progression of
  MEMS technology  \cite{Tabeling}, and tomography   \cite{Herman}, may provide a new route
  towards this goal. If
these biophysical considerations are experimentally verified, they
would   prove that trees can be an example to use technologies for
liquids under tension connected with liquids in contact with solid
substrates  at nanoscale range. They would provide a context in
which nanofluid mechanics points to a rich array of  biological
physics and future technical challenges.
\\
It is wondering to observe that the density-functional theory
expressed by a   rough energetic model   with a surface
density-functional at the walls enables to obtain a good order of
 the ascent of sap magnitude. The result is obtained without too complex
weighted density-functional and without taking   account of
quantum effects  corresponding to less than an Amgstr\"{o}m length
scale. These   observations seem  to prove that this
 kind of functionals can be a good tool to study models of
 liquids in contact with solids at a small nanoscale range.\newline
  Moreover there exists no conflict
   between thermodynamics and
 cohesion-tension theory \cite{gouin6,gouin3}.
\vskip 0.5cm \noindent
 \footnotesize \textbf{Acknowledgements:}
\emph{H.G. thanks the Accademia Nazionale dei Lincei, the Istituto
Nazionale di Alta Matematica "F. Severi" (INdAM),
 and the Gruppo Nazionale per la
Fisica Matematica (GNFM), for their nice invitation and support to
the conference on
 "New
Frontiers in Continuum Mechanics"  hold at the Academy on 21 and
22 June 2016.}

 \vskip 1cm
 \noindent {\Large

\end{document}